\documentstyle[12pt]{article}
\def\be{\begin{equation}}
\def\ee{\end{equation}}
\def\bea{\begin{eqnarray}}
\def\eea{\end{eqnarray}}
\def\bq{\begin{quote}}
\def\eq{\end{quote}}
\def\bseq{\begin{subequation}}
\def\eseq{\end{subequation}}
\def\bsea{\begin{subeqnarray}}
\def\esea{\end{subeqnarray}}
\def\simlt{\mathrel{\lower2.5pt\vbox{\lineskip=0pt\baselineskip=0pt
           \hbox{$<$}\hbox{$\sim$}}}}
\def\simgt{\mathrel{\lower2.5pt\vbox{\lineskip=0pt\baselineskip=0pt
           \hbox{$>$}\hbox{$\sim$}}}}
\def\ov{\overline}
\def\ie{i.e.~}
\def\eg{e.g.~}
\def\epem{e^+e^-}

\def\tb{\tan\beta}

\def\mpl{M_{\rm P}}
\def\mhc{m_{H^{\pm}}}
\def\msq{m_{\tilde{q}}}
\def\mglu{m_{\tilde{g}}}

\def\sell{\tilde{e}_L}
\def\selr{\tilde{e}_R}
\def\smul{\tilde{\mu}_L}
\def\smur{\tilde{\mu}_R}
\def\slepl{\tilde{l}_L}
\def\slepr{\tilde{l}_R}
\def\snu{\tilde{\nu}}
\def\snubar{\tilde{\ov{\nu}}}
\def\tev{\rm \; TeV}
\def\gev{\rm \; GeV}
\def\pb{\rm \; pb}
\def\sabsq{\sin^2 (\beta - \alpha)}
\def\cabsq{\cos^2 (\beta - \alpha)}
\begin{document}
\begin{titlepage}
\vspace*{-1cm}
\noindent
\phantom{DRAFT}
\hfill{CERN-TH.6357/91}
\vskip 2.0cm
\begin{center}
{\Large\bf The quest for low-energy supersymmetry}
\\
{\Large\bf and the role of high-energy $\epem$ colliders }
\end{center}
\vskip 1.0cm
\begin{center}
{\large F. Zwirner}\footnote{On leave from INFN, Sezione di Padova, Italy.}
\end{center}
\begin{center}
Theory Division, CERN, \\
Geneva, Switzerland \\
\end{center}
\vskip 1.0cm
\begin{abstract}
\noindent
The motivations for low-energy supersymmetry and the main features of the
minimal supersymmetric extension of the Standard Model are reviewed.
Possible non-minimal models and the issue of gauge coupling unification
are also discussed.
Theoretical results relevant for supersymmetric particle searches
at present and future accelerators are presented, with emphasis on
the role of a proposed $\epem$ collider with $\sqrt{s} = 500 \gev$.
In particular, recent results on radiative corrections to supersymmetric
Higgs boson masses and couplings are summarized, and their implications
for experimental searches are discussed in some detail.
\end{abstract}
\vskip 1.0cm
\begin{center}
{\em
Plenary talk at the Workshop on Physics and Experiments with Linear
\\
Colliders, Saariselk\"a, Lapland, Finland, 9--14 September 1991}
\end{center}
\vfill{
CERN-TH.6357/91
\newline
\noindent
December 1991}
\end{titlepage}
\setcounter{footnote}{0}
\section{Introduction}

Realistic models of low-energy supersymmetry have been studied for about
15 years, starting with the pioneering works of Fayet [\ref{fayet}] and
continuing with more and more systematic investigations [\ref{mssm}], but
there is no decisive experimental evidence yet either in favour of or
against this idea. It is then almost a duty for the theoretical speaker
on the subject (the experimental aspects are discussed in ref.~[\ref{grivaz}])
to argue in favour of the following two statements:
\begin{itemize}
\item
Low-energy supersymmetry is, today more than ever, a {\em
phenomenologically viable} and {\em theoretically motivated}
extension of the Standard Model.
\item
High-energy $\epem$ colliders can play a {\em crucial} role
in testing it experimentally.
\end{itemize}
With the above two goals in mind, the discussion will be organized as follows.
This introduction will end with a brief reminder of the motivations
for low-energy supersymmetry. Sect. 2 will introduce the Minimal
Supersymmetric extension of the Standard Model (MSSM), and its possible
non-minimal alternatives. Plausible theoretical constraints on the
MSSM, including the ones coming from gauge coupling unification, will
be also discussed. Sects. 3 and 4 will take a closer look at the particle
spectrum of the MSSM, thus providing an introduction to the experimental
discussion of ref.~[\ref{grivaz}]. The present limits from LEP~I and Tevatron
and the expected sensitivity of LEP~II and LHC/SSC will be reviewed,
followed by some theoretical considerations on the potential of a 500 GeV
$\epem$ collider (EE500). Sect. 3  will discuss recent results on radiative
corrections to Higgs boson masses and couplings, and their implications
for experimental searches. Sect. 4  will deal with supersymmetric partners
of quarks, leptons, gauge and Higgs bosons.  Finally, sect. 5 will
contain some concluding remarks.
\subsection{Motivations for low-energy supersymmetry}

There are many good reasons to believe that supersymmetry [\ref{susy}]
and its local version, supergravity [\ref{sugra}], could be relevant
in a fundamental theory of particle interactions. Symmetries, even
when broken, have been very important in establishing modern
particle theory as we know it today: supersymmetry is the
most general symmetry of the S-matrix consistent with relativistic
quantum field theory [\ref{hls}], so it is not inconceivable that Nature
might make some use of it. Also, superstrings [\ref{gsw}] are the present
best candidates for a consistent quantum theory unifying gravity with all
the other fundamental interactions, and supersymmetry appears to play a
very important role for the quantum stability of superstring solutions in
four-dimensional space-time. Experimental data, however, tell us that
supersymmetry is not realized exactly, and none of the above motivations
gives us any insight about the scale of supersymmetry breaking.

The only motivation for low-energy supersymmetry, \ie  supersymmetry
effectively broken around the electroweak scale, comes from the
{\it naturalness} or {\it hierarchy} problem [\ref{nat}] of the
Standard Model (SM), whose formulation will now be sketched. Despite its
remarkable phenomenological success [\ref{lphep}], it is impossible not to
regard the SM as an effective low-energy theory, valid up to some energy
scale $\Lambda$, at which it is replaced by some more fundamental theory.
Certainly $\Lambda$ is less than the Planck scale $\mpl \sim 10^{19}
\gev$, since one needs a theory of quantum gravity to describe physics
at these energies. However, the study of the Higgs sector of the SM
suggests that $\Lambda$ should rather be close to the Fermi scale,
$G_{\rm F}^{-1/2} \sim 300 \gev$. The argument goes as follows. Consistency of
the SM requires the SM Higgs mass to be less than $O(1 \tev)$. If one
then tries to extend the validity of the SM to energy scales $\Lambda
\gg G_{\rm F}^{-1/2}$, one is faced with the fact that in the SM there is
no symmetry to justify the smallness of the Higgs mass with respect
to the (physical) cut-off $\Lambda$. This is apparent from the fact that
in the SM one-loop radiative corrections to the Higgs mass are quadratically
divergent. Motivated by this problem, much theoretical effort has been devoted
to finding descriptions of electroweak symmetry breaking which modify the
SM at scales $\Lambda \sim G_{\rm F}^{-1/2}$. Here supersymmetry comes into
play because of its improved ultraviolet behaviour with respect to
ordinary quantum field theories [\ref{nrt}], due to cancellations
between bosonic and fermionic loop diagrams. If one wants to have
a low-energy effective Lagrangian valid up to scales $\Lambda \gg
G_{\rm F}^{-1/2}$, with one or more elementary scalar fields, kept light
without unnatural fine-tunings of parameters, the solution [\ref{les}]
is to introduce supersymmetry, effectively broken in the vicinity of the
electroweak scale. This does not yet explain why the scale $M_{\rm SUSY}$
of supersymmetry breaking is much smaller than $\Lambda$, but at least
links the Fermi scale $G_{\rm F}^{-1/2}$ to the supersymmetry-breaking scale
$M_{\rm SUSY}$, and makes the hierarchy $G_{\rm F}^{-1/2} \sim M_{\rm SUSY}
<< \Lambda$ stable against radiative corrections.

\section{The MSSM}

The most economical realization of low-energy supersymmetry is the
Minimal Supersymmetric extension of the Standard Model [\ref{mssm}],
whose defining assumptions are listed below.

\begin{itemize}
\item[1:]
{\bf Minimal gauge group.}
\\
In the MSSM, the gauge group is just $G=SU(3)_C \times SU(2)_L \times
U(1)_Y$, as in the SM.  Supersymmetry then implies that spin-1 gauge
bosons belong to {\em vector superfields}, together with their
spin-$\frac{1}{2}$ superpartners, the {\em gauginos}.
\item[2:]{\bf Minimal particle content.}
\\
The MSSM contains just three generations of quark and lepton
spin-$\frac{1}{2}$ fields, as does the SM, but embedded in {\em
chiral superfields} together with their spin-0 superpartners, the
{\em squarks} and the {\em sleptons}. In addition, to give masses to all
charged fermions and to avoid chiral anomalies, one is forced to introduce
two more chiral superfields, containing two complex spin-0 Higgs doublets
and their spin-$\frac{1}{2}$ superpartners, the {\em higgsinos}.
\item[3:]{\bf Exact $R$-parity.}
\\
Once the gauge group and the particle content are given, to determine a
globally supersymmetric Lagrangian, ${\cal L}_{\rm SUSY}$, one must specify an
analytic function of the chiral superfields, the {\em superpotential}. To
enforce baryon and lepton number conservation in renormalizable
interactions, in the MSSM one imposes a discrete, multiplicative
symmetry called $R$-parity, defined as
\be
\label{rdef}
R = (-1)^{2s+3B+L},
\ee
where $s$ is the spin quantum number. In practice, the $R$-parity
assignments are $R=+1$ for all ordinary particles (quarks, leptons,
gauge and Higgs bosons), $R=-1$ for their superpartners
(squarks, sleptons, gauginos and higgsinos). The most general
superpotential compatible with gauge invariance, renormalizability
and $R$-parity is
\be
\label{suppot}
f = h^U Q U^c H_2 + h^D Q D^c H_1
+ h^E L E^c H_1 + \mu H_1 H_2 \, ,
\ee
where $Q,U^c,D^c,L,E^c$ are the chiral superfields containing the
left-handed components of ordinary quarks and leptons, $H_1$ and $H_2$
are the two Higgs chiral superfields, and family and group indices
have been left implicit for notational simplicity.
The first three terms are nothing else than the supersymmetric
generalization of the SM Yukawa couplings, whereas the fourth one
is a globally supersymmetric Higgs mass term. Exact $R$-parity has
very important phenomenological consequences: ($R$-odd) supersymmetric
particles are always produced in pairs, their decays always involve an
odd number of supersymmetric particles in the final state, and the
lightest supersymmetric particle (LSP) is absolutely stable.
\item[4:]{\bf Soft supersymmetry breaking.}
\\
The above three assumptions are sufficient to completely determine a globally
supersymmetric renormalizable Lagrangian, ${\cal L}_{\rm SUSY}$. The MSSM
Lagrangian is obtained by adding to ${\cal L}_{\rm SUSY}$ a collection
${\cal L }_{soft}$ of explicit but {\em soft} supersymmetry-breaking
terms, which preserve the good ultraviolet properties of supersymmetric
theories. In general, ${\cal L }_{soft}$ contains [\ref{soft}] mass terms
for scalar fields and gauginos, as well as a restricted set of scalar
interaction terms proportional to the corresponding superpotential couplings
\be
\label{lsoft}
\begin{array}{ccl}
- {\cal L }_{soft}
& = &
\sum_i \tilde{m}_i^2 | \varphi_i|^2 + {1 \over 2} \sum_A M_A
\ov{\lambda}_A \lambda_A + \left( h^U A^U Q U^c H_2
\right.
\\
& + &
\left.
h^D A^D Q D^c H_1 + h^E A^E L E^c H_1 + m_3^2 H_1 H_2 + {\rm h.c.}
\right),
\end{array}
\ee
where $\varphi_i$ ($i=H_1,H_2,Q,U^c,D^c,L,E^c$) denotes the generic spin-0
field, and $\lambda_A$ ($A=1,2,3$) the generic gaugino field. Observe that,
since $A^U,A^D$ and $A^E$ are matrices in generation space, the most general
form of ${\cal L }_{soft}$ contains in principle a huge number of free
parameters. Moreover, for generic values of these parameters one
encounters phenomenological problems with flavour-changing neutral
currents [\ref{fcnc}], with new sources of CP-violation \footnote{
The phenomenology of CP violation in supersymmetric models has been discussed
recently, in connection with high-energy $\epem$ colliders, in
ref.~[\ref{cpnew}].} [\ref{cp}] and with charge- and colour-breaking vacua.
\item[5:]{\bf Unification assumptions.}
\\
All the above problems can be solved at once if one assumes that the
running MSSM parameters, defined at the one-loop level and in a
mass-independent renormalization scheme, obey a certain number
of boundary conditions at some grand-unification scale $M_U$. First of
all, one assumes grand unification of the gauge couplings
\be
\label{gcu}
g_3 (M_U) =
g_2 (M_U) =
g_1 (M_U) \equiv
g_U,
\ee
where $g_1 = \sqrt{3/5} \cdot g'$ as in most grand-unified models.
Furthermore, one assumes that all soft supersymmetry-breaking
terms can be parametrized, at the scale $M_U$, by a universal
gaugino mass
\be
\label{gmu}
M_3 (M_U) =
M_2 (M_U) =
M_1 (M_U) \equiv
m_{1/2} \, ,
\ee
a universal scalar mass
\be
\label{smu}
\tilde{m}_{H_1}^2 (M_U) =
\tilde{m}_{H_2}^2 (M_U) =
\tilde{m}_Q^2 (M_U) =
\ldots =
\tilde{m}_{E^c}^2 (M_U) \equiv
m_0^2 \, ,
\ee
and a universal trilinear scalar coupling
\be
\label{aaa}
A^U (M_U) =
A^D (M_U) =
A^E (M_U) \equiv
A \, ,
\ee
whereas $m_3^2$ remains in general an independent parameter. In addition,
all possible CP-violating phases besides the Kobayashi-Maskawa one are set
to zero at the scale $M_U$.
\end{itemize}

\subsection{Non-minimal alternatives to the MSSM}

The above assumptions, which define the MSSM, are plausible but not compulsory.
Relaxing them leads to non-minimal supersymmetric extensions of the SM,
which typically increase the number of free parameters without a
corresponding increase of physical motivation.

For example, relaxing assumption 1, a low-energy gauge group larger than the
SM one could be considered, as is possible in non-minimal grand-unification
schemes and in some string compactifications, and as was originally suggested
in some models for spontaneous breaking of global supersymmetry. However, the
present limits on the masses and mixing of extra gauge bosons are so stringent
that such a departure is certainly not motivated by now.

Similarly, there are various possibilities to enlarge the particle content
of the MSSM, relaxing assumption 2. One possibility is the introduction of
additional chiral superfields with the quantum numbers of exotic states
contained in the fundamental $\underline{27}$ representation of $E_6$: under
assumption 1, however, these states have naturally superheavy masses and
decouple from the low-energy effective theory. A particularly popular
variation, which corresponds to the simplest non-minimal model
[\ref{fayet},\ref{singlet}], is constructed by adding
a gauge-singlet Higgs superfield $N$ and by requiring purely
trilinear superpotential couplings. Without unification assumptions, this model
has already two more parameters than the MSSM, but with an assumption
analogous to eq.~(\ref{aaa}) the number of free parameters remains the same
as in the MSSM. Folklore arguments in favour of this model are that it
avoids the introduction of a supersymmetry-preserving mass parameter
$\mu \sim G_{\rm F}^{-1/2}$, and that the homogeneity properties of its
superpotential recall the structure of some superstring effective
theories. A closer look, however, shows that these statements should be taken
with a grain of salt. First, in the effective low-energy theory with softly
broken global supersymmetry, the supersymmetric mass $\mu \sim
G_{\rm F}^{-1/2}$ could well be a remnant of local supersymmetry breaking, if
the underlying supergravity theory has a suitable structure of interactions
[\ref{muproblem}]. Moreover, when embedded in a grand-unified
theory, the non-minimal model with a singlet Higgs field might develop
dangerous instabilities [\ref{instab}]. Also, the trilinear $N^3$
superpotential coupling, which is usually invoked to avoid a massless axion,
is typically absent in string models. Phenomenological aspects of the
non-minimal model with an extra singlet have been studied recently, in
connection with high-energy $\epem$ colliders, in ref.~[\ref{sinnew}],
and will not be discussed here.

Assumption 3 is of crucial importance, since relaxing it can drastically
modify the phenomenological signatures of supersymmetry. If one does not
impose $R$-parity, the most general superpotential compatible with gauge
invariance and renormalizability contains, besides the terms of eq.
(\ref{suppot}), also the following ones:
\be
\label{supbis}
\Delta f  =
\lambda Q D^c L + \lambda' L L E^c + \mu' L H_2 + \lambda'' U^c D^c D^c.
\ee
The first three terms on the right-hand side of eq. (\ref{supbis}) obey the
selection rule $\Delta B = 0, |\Delta L|=1$, and the last one the selection
rule $\Delta L = 0, |\Delta B|=1$.
Their simultaneous presence would be phenomenologically unacceptable,
since they could induce, for example, fast proton decay mediated by
$\tilde{d}^c$ squarks. However, imposing discrete symmetries weaker than
$R$-parity one can allow for some of the terms in eq. (\ref{supbis}), and
therefore for explicit $R$-parity breaking, in a phenomenologically acceptable
way [\ref{rbreak}].
Another possibility [\ref{spontr}] is that $R$-parity is spontaneously broken
by the VEV of a sneutrino field, but it is by now experimentally ruled out
by LEP data. In order to obtain acceptable models with spontaneously broken
$R$-parity,
one would need to introduce several extra fields and parameters. The
phenomenology of models with broken $R$-parity at high-energy $\epem$
colliders has been recently studied in ref.~[\ref{dreiner}], and will
not be discussed here.

To comment assumption 4, one has to discuss the problem of supersymmetry
breaking. Models with spontaneously broken global supersymmetry have to face
several phenomenological difficulties, which can be solved only at the
price of introducing rather baroque constructions. Present theoretical
ideas, however, favour the possibility that supersymmetry is spontaneously
broken in the {\em hidden sector} of some underlying supergravity (or
superstring) model, communicating with the {\em observable} sector (the one
containing the states of the MSSM) only via gravitational interactions.
As for the precise mechanism of spontaneous supersymmetry breaking, there
are several suggestions, among which non-perturbative phenomena such as
gaugino condensation [\ref{gcond}] and string constructions such as
coordinate-dependent compactifications [\ref{ss}], but none of them has
yet reached a fully satisfactory formulation. It then appears to be a sensible
choice to parametrize supersymmetry breaking in the low-energy effective
theory by a collection of soft terms, without strong assumptions on the
underlying mechanism for spontaneous supersymmetry breaking.

Besides solving naturally the phenomenological problems connected with
flavour-changing neutral currents, new sources of CP violation, charge
and colour breaking vacua, and proliferation of free parameters,
assumption 5 is strongly suggested by ideas about grand unification
and spontaneous breaking of local supersymmetry in a {\em hidden} sector;
it receives further support by the present indications on the structure
of the low-energy effective supergravity theories of string models. We
shall discuss later other phenomenological and theoretical facets of
the unification assumptions.

\subsection{Supersymmetric grand-unification}

Starting from the boundary condition of eq. (\ref{gcu}), one can solve the
appropriate renormalization group equations (RGE) to obtain the running
gauge coupling constants $g_A (Q)$ ($A=1,2,3$) at scales $Q << M_U$. At
the one-loop level, and assuming that there are no new physics thresholds
between $M_U$ and $Q$, one finds [\ref{gqw}]
\be
\label{running}
\frac{1}{g_A^2(Q)} = \frac{1}{g_U^2} +
\frac{b_A}{8 \pi^2} \log \frac{M_U}{Q}
\;\;\;\;\;
(A=1,2,3) \, ,
\ee
where the one-loop beta-function
coefficients $b_A$  depend only on the $SU(3)_C \times SU(2)_L \times
U(1)_Y$ quantum numbers of the light particle spectrum. In the MSSM
\be
\label{bsusy}
b_3 = - 3,
\;\;\;\;\;
b_2 = 1,
\;\;\;\;\;
b_1 = \frac{33}{5},
\ee
whereas in the SM
\be
\label{bsm}
b_3^0 = - 7,
\;\;\;\;\;
b_2^0 = - \frac{19}{6},
\;\;\;\;\;
b_1^0 =   \frac{41}{10}.
\ee
Starting from three input data at the electroweak scale, for example
[\ref{lphep}]
\be
\label{alfa3}
\alpha_3 (m_Z) = 0.118 \pm 0.008,
\ee
\be
\label{alfaem}
\alpha_{em}^{-1} (m_Z) = 127.9 \pm 0.2,
\ee
\be
\label{s2w}
\sin^2 \theta_W (m_Z) = 0.2327 \pm 0.0008,
\ee
where $\alpha_A=g_A^2/(4\pi)$, $\sin^2 \theta_W = g'^2 / (g^2 + g'^2)$,
$\alpha_{em}=\alpha_2 \cdot \sin^2 \theta_W$, and all running parameters
are defined in the modified minimal subtraction scheme $\ov{MS}$ [\ref{ms}],
one can perform consistency checks of the grand-unification hypothesis in
different models.

In the minimal $SU(5)$ model [\ref{gg}], and indeed in any other model
where eq. (\ref{gcu}) holds and the light-particle content is just that
of the SM (with no intermediate mass scales between $m_Z$ and $M_U$),
eqs. (\ref{running}) and (\ref{bsm}) are incompatible with experimental
data. This was first realized by noticing that the prediction $M_U \simeq
10^{14-15} \gev$, obtained by using as inputs eqs. (\ref{alfa3}) and
(\ref{alfaem}), is incompatible with experimental data on nucleon decay
[\ref{cz}]. Subsequently, also the prediction $\sin^2 \theta_W \simeq 0.21$
was shown to be in conflict with experimental data [\ref{costa}], and
this conflict became even more significant [\ref{amaldi}] after the recent
LEP precision measurements.

In the MSSM, assuming for simplicity that all supersymmetric particles
have masses of order $m_Z$, one obtains [\ref{drw}] $M_U \simeq 10^{16}
\gev$ (which increases the proton lifetime for gauge-boson-mediated
processes beyond the present experimental limits) and $\sin^2
\theta_W \simeq 0.23$. At the time of refs.~[\ref{drw}], when data were
pointing towards a significantly smaller value of $\sin^2 \theta_W$, this
was considered by some a potential phenomenological shortcoming of the MSSM.
The high degree of compatibility between data and supersymmetric grand
unification became manifest [\ref{costa}] only later, after improved data on
neutrino-nucleon deep inelastic scattering were obtained, and was recently
re-emphasized,
after the LEP precision measurements, in refs.~[\ref{grz},\ref{amaldi}].
One should not forget, however, that unification of the MSSM
is not the only solution which can fit the data of eqs.
(\ref{alfa3})--(\ref{s2w}): for example, non-supersymmetric models
with {\em ad hoc} light exotic particles or intermediate symmetry-breaking
scales [\ref{adhoc}] could also do the job. The MSSM, however,
stands out as the simplest physically motivated solution.

If one wants to make the comparison between low-energy data and
the predictions of specific grand-unified models more precise,
there are several factors that should be further taken into account.
After the inclusion of higher-loop corrections and threshold effects,
eq. (\ref{running}) is modified as follows
\be
\label{jpd}
\frac{1}{g_A^2(Q)} = \frac{1}{g_U^2} +
\frac{b_A}{8 \pi^2} \log \frac{M_U}{Q}+
\Delta_A^{th} + \Delta_A^{l>1}
\;\;\;\;\;
(A=1,2,3) \, .
\ee
In eq. (\ref{jpd}), $\Delta_A^{th}$ represents the so-called {\em threshold
effects}, which arise whenever the RGE are integrated across a particle
threshold [\ref{thr}], and  $\Delta_A^{l>1}$ represents the corrections due
to two- and higher-loop contributions to the RGE [\ref{twoloop}]. Both
$\Delta_A^{th}$ and $\Delta_A^{l>1}$ are scheme-dependent, so one should
be careful to compare data and predictions within the same renormalization
scheme. $\Delta_A^{th}$ receives contributions both from thresholds around
the electroweak scale (top quark, Higgs boson, and in SUSY-GUTs also the
additional particles of the MSSM spectrum), and from thresholds around the
grand-unification scale (superheavy gauge and Higgs bosons, and in SUSY-GUTs
also their superpartners). Needless to say, these last threshold effects
can be computed only in the framework of a specific grand-unified model,
and typically depend on a number of free parameters. Besides the effects of
gauge couplings, $\Delta_A^{l>1}$ must include also the effects
of Yukawa couplings, since, even in the simplest mass-independent
renormalization schemes, gauge and Yukawa couplings mix beyond the
one-loop order.
In minimal $SU(5)$ grand unification, and for sensible values of the
top and Higgs masses, all these corrections are small and do not affect
substantially the conclusions derived from the na\"{\i}ve one-loop analysis.
This is no longer the case, however, for supersymmetric grand unification.
First of all, one should notice that the MSSM by itself does not uniquely
define a SUSY-GUT, whereas threshold effects and even the proton lifetime
(due to a new class of diagrams [\ref{dim5}] which can be originated in
SUSY-GUTs) become strongly model-dependent. Furthermore, the simplest
SUSY-GUT [\ref{susy5}], containing only chiral Higgs superfields in the $24$,
$5$ and $\ov{5}$ representations of $SU(5)$, has a severe problem in
accounting for the huge
mass splitting between the $SU(2)$ doublets and the $SU(3)$ triplets sitting
together in the $5$ and $\ov{5}$ Higgs supermultiplets. Threshold effects
are typically larger than in ordinary GUTs, because of the much larger number
of particles in the spectrum, and in any given model they depend on several
unknown parameters. Also two-loop effects of Yukawa couplings can be
quantitatively important in SUSY-GUTs, since they depend
not only on the top-quark mass, but also on the ratio $\tb = v_2/v_1$ of
the VEVs of the two neutral Higgs fields: as will be made clearer below,
these effects become large for $m_t \simgt 140 \gev$ and $\tb \sim 1$, which
correspond to a strongly interacting top Yukawa coupling. All these effects
have been recently re-evaluated [\ref{bh}] after the enthusiasm created by
refs.~[\ref{amaldi}]. The conclusion is that, even imagining a further
reduction in the experimental errors of eqs. (\ref{alfa3})--(\ref{s2w}),
it is impossible to claim indirect evidence for supersymmetry and
to predict the MSSM spectrum with any significant accuracy.  The only
safe statement is [\ref{grz}] that, at the level of precision corresponding
to the na\"{\i}ve one-loop approximation, there is a remarkable consistency
between experimental data and the prediction of supersymmetric
grand unification, with the MSSM $R$-odd particles roughly at the
electroweak scale.

To conclude the discussion of supersymmetric grand unification, it is
worth mentioning how the unification constraints can be applied to the
low-energy effective theories of four-dimensional heterotic string models.
The basic fact to realize is that the only free parameter of these models
is the string tension, which fixes the unit of measure of the massive string
excitations. All the other scales and parameters are related to VEVs of scalar
fields, the so-called ${\em moduli}$, corresponding to flat directions of
the scalar potential. In particular, there is a relation among the string mass
$M_{\rm S} \sim \alpha'^{-1/2}$, the Planck mass $M_{\rm P} \sim
G_{\rm N}^{-1/2}$, and the unified string coupling constant $g_{\rm string}$,
which reflects unification
with gravity and implies that in any string vacuum one has one more
prediction than in ordinary field-theoretical grand unification. In a
large class of string models, one can write down an equation of the same
form as (\ref{jpd}), and compute $g_U$, $M_U$, $\Delta_A^{th}$, $\ldots$
in terms of the relevant VEVs [\ref{string}]. In the $\ov{DR}$ scheme
[\ref{drbar}], which is the most appropriate for supersymmetric models,
one finds $M_U \simeq 0.7 \times g_U \times 10^{18} \gev$, more than one
order of magnitude higher than the na\"{\i}ve extrapolations from low-energy
data illustrated before. This means that significant threshold effects are
needed in order to reconcile string unification with low-energy data:
for example, the minimal version of the flipped-$SU(5)$ model [\ref{flipped}]
is by now ruled out [\ref{lacaze}]. To get agreement, one needs some more
structure in the spectrum, either at the compactification scale or in the form
of light exotics [\ref{structure}]. Once the present string calculations
will be sufficiently generalized, unification constraints will provide
a very important phenomenological test of realistic string models.

\subsection{More on the MSSM}

It is perhaps useful, at this point of the discussion, to remind the reader
of some other phenomenological virtues and theoretical constraints of the
MSSM, besides the solution of the `technical' part of the hierarchy problem
and the grand unification of gauge couplings.

It was already said that, because of $R$-parity, the LSP is absolutely stable.
In most of the otherwise acceptable parameter space, the LSP is neutral and
weakly interacting, rarely a sneutrino and typically the lightest,
$\tilde{\chi}$, of the neutralinos
(the mass eigenstates of the neutral gaugino-higgsino sector). Then the LSP
is a natural candidate for cold dark matter [\ref{dm},\ref{ehnos}]. In
particular, for generic values of parameters one naturally avoids an excessive
$\tilde{\chi}$ relic density, but one often obtains cosmologically
interesting values for it. This should also be considered an important
consistency check of the MSSM, since a coloured or electrically charged
LSP would be in conflict with astrophysical observations [\ref{ehnos}].
Recent analyses of supersymmetric dark matter, taking into account the LEP
limits, can be found in ref.~[\ref{dmnew}].

Another remarkable fact to be noticed is that LEP precision measurements of
the $Z$ properties put little indirect constraints, via radiative
corrections [\ref{susyrad}], on the MSSM parameters. This is not the case,
for example, of technicolour and extended technicolour models, which are
severely constrained by the recent LEP data [\ref{technirad},\ref{guido}].
In the MSSM, the most important effect could be given by additional
contributions to the effective $\rho$ parameter coming from the stop-sbottom
sector: these can be sizeable only in the case of large mass splittings in
the stop-sbottom sector, in which case the upper bound on the top-quark mass,
$m_t \simlt 180 \gev$, obtained in the SM by fitting the electroweak
precision data, can be further strenghtened. However, deviations from the
SM predictions due to loops of supersymmetric particles are typically small
for generic values of the parameters.

A further predictive aspect of the MSSM is the possibility of computing
low-energy parameters, in particular the soft supersymmetry-breaking
masses, in terms of the few parameters assigned as boundary conditions at
the unification scale. To do this, it is sufficient to solve the
corresponding RGE [\ref{rge}], analogous to the ones given above
for the gauge couplings. For the gaugino masses, one finds
\be
\label{gausol}
M_A (Q) = \frac{g_A^2(Q)}{g_U^2} m_{1/2}
\;\;\;\;\;
(A=1,2,3) \, .
\ee
For the top Yukawa coupling, neglecting mixing and the
Yukawa couplings of the remaining fermions, one gets
($t \equiv \log Q$)
\be
\label{yukawa}
\frac{d h_t}{d t} =  \frac{h_t}{8 \pi^2}
\left(
- \frac{8}{3} g_3^2
- \frac{3}{2} g_2^2
- \frac{13}{18} g'^2
+ 3 h_t^2
\right).
\ee
A close look at eq. (\ref{yukawa}) can give us some important information
about the top-quark mass in the MSSM. The important thing to realize is that
the running top Yukawa coupling has an effective infrared fixed point
[\ref{bagger}], smaller than in the SM case [\ref{pendleton}].
However high the value one assigns to it at the unification scale,
$h_t$ evaluated at the electroweak scale never exceeds a
certain maximum value $h_t^{max} \simeq 1$. This implies that, for any given
value of $\tan \beta$, there is a corresponding maximum value for the top
quark mass. A naive one-loop calculation gives
\be
\label{topmax}
\begin{array}{rccccc}
\tb: & 1 & 2 & 4 & 8 & \infty \\
m_t^{max} \; (GeV): & 139 & 176 & 191 & 195 & 196 \\
\end{array}.
\ee
For the soft supersymmetry-breaking scalar masses, under the same assumptions
as above, and considering for the moment the sfermions of the third family,
one finds
\be
\label{softrge}
{d \tilde{m}_i^2 \over {d t}} =   {1 \over {8 \pi^2}}
\left[
- \sum_{A=1,2,3} c_A(i) g_A^2 M_A^2
+ c_t (i) h_t^2 F_t
\right],
\ee
where $i=H_1,H_2,Q,U^c,D^c,L,E^c$,
\be
F_t  \equiv  \tilde{m}_Q^2 + \tilde{m}_{U^c}^2 + \tilde{m}_{H_2}^2 + A_t^2,
\ee
and the $c_A(i),c_t(i)$ coefficients are given by
\begin{equation}
\label{coeff}
\begin{array}{rccccccc}
i: & H_1 & H_2 & Q & U^c & D^c & L & E^c \\
c_3(i): & 0 & 0 & \frac{16}{3} & \frac{16}{3} & \frac{16}{3} & 0 & 0 \\
c_2(i): & 3 & 3 & 3 & 0 & 0 & 3 & 0 \\
c_1(i): & \frac{3}{5} & \frac{3}{5} & \frac{1}{15} & \frac{16}{15} &
\frac{4}{15} & \frac{3}{5} & \frac{12}{5} \\
c_t(i): & 0 & 3 & 1 & 2 & 0 & 0 & 0
\end{array}
\ee
Similar equations can be derived for the remaining soft supersymmetry-breaking
parameters and for the superpotential Higgs mass $\mu$. Also, the inclusion of
the complete set of Yukawa couplings, including mixing, is straightforward.
In general, the RGEs for superpotential couplings and soft
supersymmetry-breaking parameters have to be solved by numerical methods.
Analytical
solutions can be obtained for the soft squark and slepton masses when
the corresponding Yukawa couplings are negligible:
\begin{equation}
\label{softsol}
\tilde{m}_i^2 = m_0^2 + m_{1/2}^2
\sum_{A=1}^{3} \frac{c_A(i)}{2 b_A} \left(1 - \frac{1}{F_A^2} \right),
\end{equation}
where
\begin{equation}
F_A = 1 + \frac{b_A}{8 \pi^2} g_U^2 \log \frac{M_U}{Q} \, .
\end{equation}
For example, one gets $\tilde{m}_Q^2, \tilde{m}_{U^c}^2,
\tilde{m}_{D^c}^2 \sim m_0^2 + (5 - 8) \, m_{1/2}^2$,
$\tilde{m}_L^2 \sim m_0^2 + 0.5 \, m_{1/2}^2$,
$\tilde{m}_{E^c}^2 \sim m_0^2 + 0.15  \, m_{1/2}^2$.
It should be stressed that also eqs. (\ref{gausol}) and (\ref{softsol}),
in analogy with eq. (\ref{running}), are valid up to higher-order corrections
and threshold effects, so their accuracy should not be overestimated.

One of the most attractive features of the MSSM is the possibility
of describing the spontaneous breaking of the electroweak gauge
symmetry as an effect of radiative corrections [\ref{radbre}],
via a generalization of the mechanism discussed first by Coleman
and E. Weinberg [\ref{cw}] in the context of the SM. It is remarkable
that, starting from universal boundary conditions at the unification
scale, it is possible to explain naturally why fields carrying colour
or electric charge do not acquire non-vanishing VEVs, whereas the neutral
components of the Higgs doublets do. We give here a simplified description
of the mechanism in which the physical content is transparent, and we
comment later on the importance of a more refined treatment. The starting
point is a set of boundary conditions on the independent model parameters
at the unification scale $Q = M_U$. One then evolves all the running
parameters from the grand-unification scale to a low scale $Q \sim m_Z$,
according to the RGE, and considers the renormalization-group-improved
tree-level potential
\be
\label{v0}
\begin{array}{ccl}
V_0 (Q) & = &
m_1^2 \left| H_1 \right|^2 +
m_2^2 \left| H_2 \right|^2 +
m_3^2 \left( H_1 H_2 + {\rm h.c.} \right)
\\ & & \\ & + &
\frac{1}{8}g^2 \left( H_2^{\dagger} {\vec\sigma} H_2
+ H_1^{\dagger} {\vec\sigma} H_1 \right)^2 +
\frac{1}{8}g'^2 \left(  \left| H_2 \right|^2 -
\left| H_1 \right|^2 \right)^2,
\end{array}
\ee
where
\be
m_1^2 \equiv \tilde{m}_{H_1}^2 + \mu^2,
\;\;\;\;\;
m_2^2 \equiv \tilde{m}_{H_2}^2 + \mu^2,
\ee
and it is not restrictive to choose a field basis such that $m_3^2 \leq 0$.
All masses and coupling constants in $V_0(Q)$ are running parameters,
evaluated at the scale $Q$. The minimization of the potential in eq.
(\ref{v0}) is straightforward. To generate non-vanishing VEVs $v_1 \equiv
\langle H_1^0 \rangle$ and $v_2 \equiv \langle H_2^0 \rangle$, one needs
\be
{\cal B} \equiv m_1^2 m_2^2 - m_3^4 < 0 \, .
\ee
In addition, a certain number of conditions have to be satisfied
to have a stable minimum with the correct amount of symmetry
breaking and with unbroken colour, electric charge, baryon and lepton
number: for example, all the running squark and slepton masses have to be
positive. A crucial role in the whole process is played by the top
Yukawa coupling, which strongly influences the RGE for $\tilde{m}_{H_2}^2$,
as should be clear from eqs. (\ref{softrge})--(\ref{coeff}). For appropriate
boundary conditions, the RGE drive ${\cal B} < 0$ at scales $Q \sim m_Z$,
whereas all the squark and slepton masses remain positive as desired, to
give a phenomenologically acceptable breaking of the electroweak symmetry.

The use of $V_0(Q)$ is very practical for a qualitative discussion as the
one given above, but it relies on the assumption that, once the leading
logarithms have been included in the running parameters, all the remaining
one-loop corrections to the scalar potential can be neglected at the scale
$Q \sim m_Z$. However, as shown for example in ref.~[\ref{grz}], for a
quantitative discussion of gauge symmetry breaking it is necessary to use
the full one-loop effective potential, which in the Landau gauge and in the
$\ov{DR}$ renormalization scheme [\ref{drbar}] is given by
\begin{equation}
\label{v1}
V_1 (Q) = V_0 (Q) + \frac{1}{64 \pi^2} \; {\rm Str} \; \left\{
{\cal M}^4(Q) \left[ \log \frac{{\cal M}^2(Q)}{Q^2} - \frac{3}{2}
\right] \right\}.
\end{equation}
In eq. (\ref{v1}), ${\rm Str} \; f({\cal M}^2) = \sum_i (-1)^{2J_i} (2 J_i +
1 ) f(m_i^2)$ denotes the conventional supertrace, where $m_i^2$ is the
field-dependent mass eigenvalue of the i-th particle of spin $J_i$, and
field-independent terms have been neglected. To give an example, the
VEVs determined from $V_0(Q)$ are strongly scale-dependent, whereas
the ones determined from $V_1(Q)$ are not, as it should be. Only at
a scale $\hat{Q}$, of the order of the stop masses, is the use of $V_0(Q)$
a good approximation. This is a result of the fact that mass-independent
renormalization schemes, like $\ov{MS}$ or $\ov{DR}$, do not
automatically include decoupling: since the most important contributions
to $V_1(Q)$ come from the stop sector, the optimal scale at which
to freeze the evolution of the running parameters turns out to be
of the order of the stop masses. Another aspect of this effect,
with important phenomenological consequences, are the radiative corrections
to Higgs boson masses and couplings, which will be discussed in the following
section.

To conclude the discussion of radiative symmetry breaking, we
show now that in the MSSM, with universal boundary conditions,
one expects
\begin{equation}
\label{brange}
1 \simlt \tan \beta \simlt \frac{m_t}{m_b} \, .
\end{equation}
The simplest proof relies on the relation, derived from the minimization
of $V_0(Q)$
\begin{equation}
\label{tanbeta}
\frac{v_2}{v_1}
=
\frac{m_1^2 + m_Z^2 / 2}{m_2^2 + m_Z^2 / 2} \, .
\end{equation}
The boundary conditions at the unification scale is $m_1^2(M_U)=m_2^2(M_U)$,
and, neglecting as before all Yukawa couplings except $h_t$, the RGE for the
difference $m_1^2-m_2^2$ reads
\begin{equation}
\label{diff}
\frac{d (m_1^2 - m_2^2)}{d t} = - \frac{3}{8 \pi^2} h_t^2 F_t \, .
\end{equation}
Imagine now that $\tan \beta < 1$, and observe that the top and bottom masses
are given by $m_t^2 = h_t^2 v_2^2$ and $m_b^2 = h_b^2 v_1^2$, respectively.
Then $m_t \gg m_b$ implies $h_t \gg h_b$, which makes eq.
(\ref{diff}) a good approximation. Solving now eq. (\ref{diff}) at the
scale $\hat{Q}$, where the use of $V_0(Q)$ is justified, and observing that
it is always $F_t > 0$, one finds $m_1^2 > m_2^2$. But eq. (\ref{tanbeta})
then tells us that $\tan \beta > 1$, in  contradiction with the starting
assumption. Similarly, including in eq. (\ref{diff}) the contributions of
the bottom and $\tau$ Yukawa couplings, one can prove that $\tan \beta
\simlt m_t / m_b$.

\section{Higgs bosons}

We begin the discussion of the MSSM particle spectrum with the ($R$-even)
Higgs boson sector. As explained in the previous section, the MSSM contains
two complex Higgs doublets of opposite hypercharge, $H_1 \equiv (H_1^0,
H_1^-)$ and $H_2 \equiv (H_2^+, H_2^0)$. After their neutral components
develop non-vanishing VEVs, $v_1$ and $v_2$, which can be taken to be real
and positive without loss of generality, one is left with five physical
degrees of freedom. Three of these are neutral (two CP-even, $h$ and $H$,
and one CP-odd, $A$) and two are charged ($H^{\pm}$). The starting point
for a discussion of Higgs-boson masses and couplings in the MSSM is the
potential of eq. (\ref{v0}). Besides the minimization conditions, which
relate $v_1$ and $v_2$ with the potential parameters, a physical constraint
comes from the fact that the combination $(v_1^2+v_2^2)$, which determines
the $W$- and $Z$-boson masses, must reproduce their measured values. Once this
constraint is imposed, in the approximation of eq.~(\ref{v0}) the MSSM Higgs
sector
contains only two independent parameters. A convenient choice, which will be
adopted here, is to take as independent parameters $m_A$, the physical mass
of the CP-odd neutral boson, and $\tan \beta \equiv v_2 / v_1$. The parameter
$m_A$ is essentially unconstrained, even if naturalness arguments suggest
that it should be smaller than $O(500 \gev)$, whereas for $\tan \beta$ the
range permitted in the MSSM is given by formula (\ref{brange}).

In the approximation of eq. (\ref{v0}), the mass matrix of neutral CP-even
Higgs bosons reads
\be
\label{cpeven0}
\left( {\cal M}_R^0 \right)^2 =
\left[
\pmatrix{
\cot\beta & -1 \cr
-1 & \tan\beta \cr}
{m_Z^2 \over 2}
+
\pmatrix{
\tan\beta & -1 \cr
-1 & \cot\beta \cr}
{m_A^2 \over 2}
\right]
\sin 2\beta
\ee
and the charged Higgs mass is given by
\be
\label{mch0}
m_{H^\pm}^2 = m_W^2 + m_A^2 \, .
\ee
{}From eq. (\ref{cpeven0}), one obtains
\be
\label{mh0}
m_{h,H}^2 = {1 \over 2} \left[
m_A^2 + m_Z^2 \mp \sqrt{(m_A^2 + m_Z^2)^2
- 4 m_A^2 m_Z^2 \cos^2 2 \beta}
\right],
\ee
and also celebrated inequalities as $m_W, m_A < m_{H^{\pm}}$,
$m_h <  m_Z < m_H$, $m_h <  m_A < m_H$. Similarly, one can easily compute all
the Higgs-boson couplings by observing that the mixing-angle $\alpha$, required
to diagonalize the mass matrix (\ref{cpeven0}), is given by
\be
\cos 2 \alpha = - \cos 2 \beta \; { { m_A^2  - m_Z^2 }\over {
 m_H^2 - m_h^2  }}
\;\;\;\;\;
\left( - {\pi \over 2} < \alpha\  {\leq}\  0 \right).
\ee
For example, the tree-level couplings of the neutral Higgs bosons
are easily obtained from the standard model Higgs couplings if one
multiplies them by some appropriate $\alpha$- and $\beta$-dependent
factors [\ref{hunter}]. An important
consequence of the structure of the Higgs potential (\ref{v0}) is the
existence of at least one neutral CP-even Higgs boson, $h$, weighing
less than $m_Z$. This raised the hope that the crucial experiment
on the MSSM Higgs sector could be entirely performed at LEP~II (with
sufficient centre-of-mass energy, luminosity and b-tagging efficiency),
and took some interest away from higher energy colliders.
However, it was recently pointed out [\ref{jap1}--\ref{hh}] that
the Higgs-boson masses are subject to large radiative corrections,
associated with the top quark and its $SU(2)$ and supersymmetric
partners\footnote{Previous studies [\ref{lisher}--\ref{berger}] either
neglected the case of a heavy top quark [\ref{lisher},\ref{gt}], or
concentrated on the violations of the neutral Higgs-mass sum rule
[\ref{berger}] without computing corrections to individual Higgs masses.}.
Several papers [\ref{bfc}--\ref{b2}] have subsequently investigated
various aspects of these corrections and their implications for experimental
searches at LEP and LHC-SSC. In the following subsection, we shall summarize
the main effects of radiative corrections on Higgs-boson parameters.

\subsection{Radiative corrections to Higgs boson-masses and couplings}

As far as Higgs-boson masses and self-couplings are concerned, a convenient
approximate way of parametrizing one-loop radiative corrections is to
substitute the tree-level Higgs potential of eq.~(\ref{v0}) with the
one-loop effective potential of eq. (\ref{v1}), and to identify masses and
self-couplings with the appropriate combinations of derivatives evaluated
at the minimum. The comparison with explicit diagrammatic calculations
[\ref{hh},\ref{yam},\ref{cpr},\ref{b},\ref{b2}] shows that the effective
potential
approximation is more than adequate for our purposes. Also, inspection shows
that the most important contributions to eq. (\ref{v1}) come from the
field-dependent mass matrices of the top and bottom quarks and squarks, whose
explicit expressions depend on a number of parameters and can be found
in ref.~[\ref{berz}]. To simplify the discussion, in the following we will
take a universal soft supersymmetry-breaking squark mass, $\tilde{m}_Q^2 =
\tilde{m}_{U^c}^2 = \tilde{m}_{D^c}^2 \equiv \msq^2$, and we will assume
negligible mixing in the stop and sbottom mass matrices, $A_t = A_b = \mu = 0$.
More complete formulae for arbitrary values of the parameters can be found in
refs.~[\ref{erz2},\ref{berz}], but the qualitative features corresponding
to the above parameter choices are representative of a very large region of
parameter space. In the case under consideration, and neglecting D-term
contributions to the squark masses, the neutral CP-even mass matrix is
modified at one loop as follows
\be
\label{cpeven1}
{\cal M}_R^2 = \left( {\cal M}_R^0 \right)^2
+
\pmatrix{
\Delta_1^2 & 0 \cr
0 & \Delta_2^2 \cr},
\ee
where
\be
\label{delta1}
\Delta_1^2 = {3 g^2 m_b^4 \over 16 \pi^2 m_W^2 \cos^2\beta}
\log\frac{m_{\tilde{b}_1}^2 m_{\tilde{b}_2}^2}{m_b^4},
\ee
\be
\label{delta2}
\Delta_2^2= {3 g^2 m_t^4 \over 16 \pi^2 m_W^2 \sin^2\beta}
\log\frac{m_{\tilde{t}_1}^2 m_{\tilde{t}_2}^2}{m_t^4}.
\ee
{}From the above expressions one can easily derive the one-loop-corrected
eigenvalues $m_h$ and $m_H$, as well as the mixing angle $\alpha$ associated
with the one-loop-corrected mass matrix (\ref{cpeven1}).
The most striking fact in eqs. (\ref{cpeven1})--(\ref{delta2}) is that the
correction $\Delta_2^2$ is proportional to $(m_t^4/m_W^2)$. This implies
that, for $m_t$ in the range allowed by experimental limits and by eq.
(\ref{topmax}), the tree-level predictions for $m_h$ and $m_H$ can be badly
violated, and so for the related inequalities. The other free parameter
is $\msq$, but the dependence on it is much milder. In the following, when
making numerical examples, we shall always choose the representative value
$\msq = 1 \tev$. The reader can easily rescale the displayed results to
different values of $\msq$. To illustrate the impact of these results, we
display in fig.~1 [\ref{kznew}]
contours of the maximum allowed value of $m_h$ (reached
for $m_A \to \infty$), in the $(m_t,\tb)$ plane. To plot different quantities
of physical interest in the $(m_A,\tb)$ plane, which is going to be the stage
of the following phenomenological discussion, one needs to fix also the value
of $m_t$. In the following, we shall work with the representative value
$m_t=140 \gev$, which is near the centre of the presently allowed range. As
an example, we show in fig.~2 contours of constant $m_h$ and  $m_H$ in the
$(m_A,\tb)$ plane. One-loop corrections to the charged Higgs mass
have also been computed in refs.~[\ref{berz}--\ref{drees}], and found to be
small, at most a few GeV, for generic values of the parameters.

The effective potential method allows also the computation of the leading
corrections to the trilinear and quadrilinear Higgs self-couplings. For
example, the
leading radiative correction to the trilinear $hAA$ coupling, which plays a
major role in the determination of the $h$ branching ratios, is [\ref{berz}]
\begin{equation}
\label{hhh1}
\lambda_{hAA}
=
\lambda_{hAA}^0
+
\Delta \lambda_{hAA} \, ,
\end{equation}
where
\begin{equation}
\label{lhaa0}
\lambda_{hAA}^0
=
- {i g m_Z \over {2 \cos \theta_W}}
\cos 2 \beta \sin (\beta + \alpha) \, ,
\end{equation}
and, neglecting the bottom Yukawa coupling and the D-term contributions
to squark masses
\begin{equation}
\label{deltahaa}
\Delta \lambda_{hAA}
=
- {i g m_Z \over {2 \cos \theta_W}}
  {3 g^2 \cos^2 \theta_W \over {8 \pi^2}}
  {\cos  \alpha \cos^2 \beta \over  {\sin^3 \beta}}
  {m_t^4 \over m_W^4}
  \log
  {\msq^2 + m_t^2 \over m_t^2} \, .
\end{equation}
Similarly, also the other Higgs self-couplings receive large corrections
$O(m_t^4/m_W^4)$.

Finally, one should consider Higgs couplings to vector bosons and fermions.
Tree-level couplings to vector bosons are expressed in terms of gauge
couplings and of the angles $\beta$ and $\alpha$. The most important part
of the radiative corrections is taken into account by using
one-loop-corrected formulae to determine $\alpha$ from the input parameters.
Other corrections are at most $O(m_t^2 / m_W^2)$ and can be
safely neglected for our purposes. Tree-level couplings to fermions are
expressed in terms of the fermion masses and of the angles $\beta$ and
$\alpha$. In this case, the leading radiative corrections can be taken
into account by using the one-loop-corrected expression for $\alpha$ and
running fermion masses, evaluated at the scale $Q$ which characterizes
the process under consideration.

\subsection{The discovery potential of LEP and LHC-SSC}

In this section, we briefly summarize the implications of the previous
results on MSSM Higgs-boson searches at LEP [\ref{bf},\ref{erz2},\ref{berz}]
and the LHC-SSC [\ref{kznew},\ref{clones}].

As already clear from tree-level analyses [\ref{hunter}], the relevant
processes for MSSM Higgs boson searches at LEP~I are $Z \to h Z^*$ and
$Z \to h A$, which play a complementary role since their rates are
proportional to $\sabsq$ and $\cabsq$, respectively. An important effect
of radiative corrections [\ref{berz}] is to render possible, for some values
of the parameters, the decay $h \to AA$, which would be kinematically forbidden
according to tree-level formulae. Experimental limits which take radiative
corrections into account have by now been obtained by the four LEP
collaborations [\ref{lep}], using different methods to present and
analyse the data, and different ranges of parameters in the evaluation of
radiative corrections. The presently excluded region of the $(m_A,\tb)$ plane,
for our standard parameter choice, is given in fig.~3 [\ref{kznew}], where
the solid line corresponds to the exclusion contour given in the first of
refs.~[\ref{lep}].

The situation in which the impact of radiative corrections is most
dramatic is the search for MSSM Higgs bosons at LEP~II. At the time
when only tree-level formulae were available, there was hope that LEP
could completely test the MSSM Higgs sector. According to tree-level formulae,
in fact, there should always be a CP-even Higgs boson with mass smaller than
($h$) or very close to ($H$) $m_Z$, and significantly coupled to the $Z$ boson.
However, as should be clear from the previous section, this result can be
completely upset by radiative corrections. A detailed evaluation of the
LEP~II discovery potential can be made only if crucial theoretical parameters,
such as the top-quark mass and the various soft supersymmetry-breaking masses,
and experimental parameters, such as the centre-of-mass energy, the luminosity,
and the b-tagging efficiency, are specified. Taking for example $\sqrt{s} =
190 \gev$, $m_t \simgt 110 \gev$, and our standard values for the soft
supersymmetry-breaking parameters, in the region of $\tan \beta$ significantly
greater than 1 the associated production of a $Z$ and a CP-even Higgs can be
pushed beyond the kinematical limit. Associated $hA$ production could be a
useful complementary signal, but obviously only for $m_h+m_A< \sqrt{s}$.
Associated $HA$ production is typically negligible at these energies.
To give a measure of the LEP~II sensitivity, we plot in fig.~3 contours
associated to two benchmark values of the total cross-section $\sigma(
e^+ e^- \to hZ, HZ, hA, HA)$. The dashed lines correspond to $\sigma = 0.2
\, {\rm pb}$ at $\sqrt{s} = 175 \gev$, which could be seen as a rather
conservative estimate of the LEP~II sensitivity. The dash-dotted lines
correspond to $\sigma = 0.05 \, {\rm pb}$ at $\sqrt{s} = 190 \gev$, which
could be seen as a rather optimistic estimate of the LEP~II sensitivity.
Of course, one should keep in mind that there is, at least in principle,
the possibility of further extending the maximum LEP energy up to values
as high as $\sqrt{s} \simeq 230-240 \gev$, at the price of introducing more
(and more performing) superconducting cavities into the LEP tunnel
[\ref{treille}].

Similar considerations can be made for charged Higgs searches at LEP~II
with $\sqrt{s} \simlt 190 \; {\rm GeV}$. In view of the $\beta^3$ threshold
factor in $\sigma ( e^+ e^- \to H^+ H^-)$, and of the large background from
$e^+ e^- \to W^+ W^-$, it will be difficult to find the $H^{\pm}$ at LEP~II
unless $\mhc \simlt m_W$, and certainly impossible unless $\mhc < \sqrt{s}/2$.
We also know [\ref{berz}--\ref{drees}] that for generic values of the
parameters there are no large negative radiative corrections to the charged
Higgs mass formula, eq. (\ref{mch0}). Thus there is very little hope of
finding the charged Higgs boson of the MSSM at LEP~II (or, to put it
differently, the discovery of a charged Higgs boson at LEP~II would most
probably rule out the MSSM).

The next question is then whether the LHC and SSC can explore the full
parameter
space of the MSSM Higgs bosons. A systematic study of this problem, including
radiative corrections, has been recently performed in ref.~[\ref{kznew}]
(see also [\ref{clones}]), following the strategy outlined in ref.~[\ref{kz}].
The analysis is complicated by the fact that the $R$-odd particles could play
a role both in the production (via loop diagrams) and in the decay (via loop
diagrams and as final states) of the MSSM Higgs bosons. For simplicity, one
can concentrate on the most conservative case in which all $R$-odd particles
are heavy enough not to play any significant role. Still, one has to perform
a separate analysis for each $(m_A,\tb)$ point, to include radiative
corrections (depending on the parameters of the top-stop-bottom-sbottom
system), and to consider Higgs boson decays involving other Higgs bosons.
We make here only a few general remarks on the LHC case, for the
representative parameter choice $m_t = 140 \gev$, $\msq = 1 \tev$,
$A_t=A_b=\mu=0$, sending the reader to ref.~[\ref{kznew}] for a more
complete discussion, and to ref.~[\ref{pauss}] for a review of recent
simulation work.

Beginning with the neutral states, when $h$ or $H$ are in the
intermediate mass range (80--130 GeV) and have approximately SM
couplings, the best prospects for detection are offered, as in the
SM, by their $\gamma \gamma$ decay mode. In general, however,
$\sigma \cdot BR (h,H \to \gamma \gamma)$ is smaller than for a
SM Higgs boson of the same mass. As a rather optimistic estimate
of the possible LHC sensitivity, we display, in fig.~4, lines
corresponding to $\sigma \cdot BR (h,H \to \gamma \gamma) \sim 2/5$
of the corresponding value for a SM Higgs of $100 \gev$. Only in the
region of the $(m_A,\tb)$ plane to the right of the line denoted by `a'
(in the case of $h$) and above the line denoted by `b' (in the case of $H$)
the $\gamma \gamma$ signal exceeds the chosen reference value.
Almost identical considerations can be made for the production of
$h$ or $H$, decaying into $\gamma \gamma$, in association with a
$W$ boson or with a $t \ov{t}$ pair.
When $H$ and $A$ are heavy, in general one cannot rely on the
$ZZ \to 4 l^{\pm}$ $(l=e,\mu)$ decay mode, which gives the
`gold-plated' Higgs signature in the SM case, since $H$ and $A$
couplings to vector-boson pairs are strongly suppressed: only for
small $\tan \beta$ and $2 m_Z \simlt m_H \simlt 2 m_t$ might the decay mode
$H \rightarrow Z Z \rightarrow 4 l^{\pm}$ still be viable despite
the suppressed branching ratio. Again, as an estimate of the possible
LHC sensitivity, we show in fig.~4, under the line denoted
by `c', the region of the $(m_A,\tb)$ plane corresponding to
$\sigma \cdot BR (H \to 4 l^{\pm}) > 10^{-3} \pb$ ($l=e,\mu$).
For very large values of $\tan \beta$, and moderately large $m_A$,
the unsuppressed decays $H,A \rightarrow \tau^+ \tau^-$ could give
visible signals, in contrast to the SM case. As a very optimistic
estimate (especially in the small $m_A$ region!)  we show in fig.~4,
above the line denoted by `d', the region of the parameter space
corresponding to $\sigma \cdot BR (H,A \to \tau^+ \tau^-) > 1 \pb$.
Finally, in the region of parameter space corresponding to $m_A \simlt m_Z$,
the charged Higgs could be discovered via the decay chain
$t \to b H^+ \to b \tau^+ \nu_{\tau}$, which competes with the standard
channel $t \to b W^+ \to b l^{\pm} \nu_l$ ($l=e,\mu,\tau$).
A convenient parameter
is the ratio $R \equiv BR(t \to \tau^+ \nu_{\tau} b)/BR(t \to \mu^+ \nu_{\mu}
b)$. As a very optimistic estimate of the LHC sensitivity, the line
of fig.~4 denoted by `e' delimits from the right the region of the $(m_A,
\tb)$ plane corresponding to $R>1.1$. For all processes considered above,
similar remarks apply also to the SSC.

In summary, a global look at figs.~3 and 4 shows that there is a high degree
of complementarity between the regions of parameter space accessible to LEP~II
and to the LHC-SSC. However, for our representative choice of parameters,
there is a non-negligible region of the $(m_A,\tb)$ plane that is presumably
beyond the reach of
both LEP~II and the LHC-SSC. This potential problem could be solved, as we
said before, by a further increase of the LEP~II energy beyond the reference
value $\sqrt{s} \simlt 190 \gev$. Otherwise, this is the ideal case for
a 500 GeV (or even less) $\epem$ collider, as we shall see below.
Even if in the future a Higgs boson will be found at LEP or the LHC-SSC,
with properties compatible with those of a MSSM Higgs boson,
it appears difficult to search effectively for all the Higgs states
of the MSSM at the above machines. Again, as we shall see below,
EE500 could play an important role in investigating the properties
of the newly discovered Higgs boson and in looking for the remaining
states of the MSSM.

\subsection{Production mechanisms at high-energy $\epem$ colliders}

We now present, following ref.~[\ref{ee500}], cross-sections for the main
production mechanisms of neutral susy Higgses in $e^+e^-$ collisions at
$\sqrt{s} = 500$ GeV, namely:
\begin{displaymath}
\begin{array}{ll}
e^+e^- \rightarrow h Z   & [\sigma \propto \sin^2(\beta-\alpha)] \, ,
\\
e^+e^- \rightarrow H Z   &  [\sigma \propto \cos^2(\beta-\alpha)] \, ,
\\
e^+e^- \rightarrow h A   &  [\sigma \propto \cos^2(\beta-\alpha)] \, ,
\\
e^+e^- \rightarrow  H A  &  [\sigma \propto \sin^2(\beta-\alpha)] \, ,
\\
e^+e^- \rightarrow h \nu \overline{\nu} &
[\sigma \propto \sin^2(\beta-\alpha)] \, ,
\\
 e^+e^- \rightarrow H \nu \overline{\nu} &
 [\sigma \propto \cos^2(\beta-\alpha)] \, ,
\\
 e^+e^- \rightarrow h e^+e^-  &  [\sigma \propto \sin^2(\beta-\alpha)]
\, ,
\\
 e^+e^- \rightarrow H e^+e^-  &  [\sigma \propto \cos^2(\beta-\alpha)]
\, .
\end{array}
\end{displaymath}
Other production mechanisms of interest are discussed in
refs.~[\ref{ee500},\ref{haber}], and details about experimental searches
can be found in refs.~[\ref{janot},\ref{koma}]. We have included radiative
corrections to the masses $m_h$, $m_H$ and to the mixing angle $\alpha$ for
our standard parameter choice. We have neglected loops from the
gauge-gaugino-Higgs-higgsino sector, which are known to give corrections
smaller than the ones we have included. We have also neglected proper
vertex corrections to vector boson-Higgs boson couplings and initial-state
radiation.

In discussing our results, it is useful to estimate the
cross-section for which we believe that any of the listed processes
will be detectable. A cross-section of $0.01$~pb will lead to
25 events for an integrated luminosity of $10$~fb$^{-1}$ after
multiplying by an efficiency of 25\%; the latter is a crude estimate
of the impact of detector efficiencies, cuts, and branching ratios
to usable decay channels. It will be helpful to keep this benchmark
cross-section value in mind as a rough criterion for where
in parameter space a particular reaction can be useful.

Fig.~5 shows contours of $\sigma (  e^+e^- \rightarrow h Z)$ and
$\sigma (  e^+e^- \rightarrow H Z)$, respectively, in the $(m_A,\tan\beta)$
plane. Observe that the two processes are truly complementary, in the sense
that everywhere in the $(m_A,\tan\beta)$ plane there is a substantial
cross-section for at least one of them ($\sigma > 0.01$ pb). This should be
an excellent starting point for experimental searches.
Similar considerations hold for $h A$, $H A$ production, whose cross-sections
are shown in fig.~6. As long as one of the two channels is kinematically
accessible, the inclusive cross-section is large enough to provide a
substantial event rate. Even in this case the two processes are complementary,
and together should be able to probe the region of parameter space
corresponding to $m_A \simlt 200$ GeV.
We now move to single Higgs production via vector-boson fusion.
The cross-sections for $h$,$H$ production via $WW$ and $ZZ$ fusion
are given in figs.~7 and 8, respectively: they have been obtained using
exact analytical formulae, rescaled from ref.~[\ref{amp}]. Obviously,
since the $AWW$ and $AZZ$ vertices
are absent at tree level, one cannot get substantial $A$ production with
this mechanism for sensible values of the parameters. The $ZZ$ fusion
processes are suppressed by an order of magnitude with respect to the
$WW$ fusion ones, but could still be useful for experimental searches.

The global picture which emerges from our results is the following. If
no neutral Higgs boson is previously discovered, at EE500 one must find
at least one neutral susy Higgs, otherwise the MSSM is ruled out. If $m_A$
is not too large, at EE500 there is the possibility of discovering
all of the Higgs states of the MSSM via a variety of processes, including
charged-Higgs-boson production, which has not been discussed here.
In the event that a neutral Higgs boson is discovered previously at LEP
or the LHC-SSC, with properties compatible with one of the MSSM Higgs
states, EE500 would still be a very useful instrument to investigate in
detail the spectroscopy of the Higgs sector, for example to distinguish
between the SM, the MSSM and possibly other non-minimal supersymmetric
extensions.

\section{$R$-odd particles}

We now briefly review the $R$-odd spectrum of the MSSM, to introduce the
discussion of supersymmetric particle searches at $\epem$ and hadron colliders.

In the spin-0 sector, one has sleptons and squarks, $\tilde{f} \equiv
( \tilde{\nu}_L, \tilde{e}_L, \tilde{e^c}_L \equiv \tilde{e}_R^*,
\tilde{u}_L, \tilde{u^c}_L \equiv \tilde{u}_R^*,
\tilde{d}_L$, $
\tilde{d^c}_L \equiv \tilde{d}_R^*)$, with generation indices
left implicit as usual. Neglecting intergenerational mixing, their diagonal
mass terms are given by
\be
\label{diagonal}
m_{\tilde{f}}^2 = \tilde{m}^2 + m_f^2 + m_D^2 \, ,
\ee
where $\tilde{m}$ is the soft supersymmetry-breaking mass, $m_f$ is
the corresponding fermion mass, and
\be
\label{dterm}
m_D^2 = m_Z^2 {\tan^2 \beta - 1 \over {\tan^2 \beta + 1}}
( Y \sin^2 \theta_W - T_{3L} \cos^2 \theta_W ) \, .
\ee
For the sfermions of the first two generations, $\tilde{f}_L$-$\tilde{f}_R$
mixing is negligible and the soft masses are given by eq. (\ref{softsol}),
so one can express $m_{\tilde{f}}$ in terms of the basic parameters $m_{1/2},
m_0$ and $\tb$. Notice for example that, neglecting the lepton masses,
$SU(2)$ invariance alone requires $m_{\tilde{\nu}}^2 = m_{\tilde{e}_L}^2
- m_W^2  [(\tan^2 \beta - 1)/(\tan^2 \beta + 1)]$. For the sfermions of the
third generation, the off-diagonal term in the $\tilde{f}_L$-$\tilde{f}_R$
mass matrix
\be
\label{offdiag}
m^2_{\tilde{f}_{LR}} = m_f \times
\left\{ \begin{array}{ll}
A_f + \mu \cdot \tb & (f=e,d) \\
A_f + \mu /     \tb & (f=u)
\end{array}
\right. ,
\ee
might be non-negligible, so that the mass eigenstates $(\tilde{f}_1,
\tilde{f}_2)$ are non-trivial admixtures of the interaction eigenstates
$(\tilde{f}_L,\tilde{f}_R)$. Also, to compute the soft contributions to
the masses in terms of the basic parameters one has to solve numerically
the associated RGE.

In the spin-$\frac{1}{2}$ sector, one has the strongly interacting gluinos,
$\tilde{g}$, with mass $m_{\tilde{g}} \equiv M_3$ directly given by eq.
(\ref{gausol}). In addition, one has the weakly interacting {\em charginos}
and {\em neutralinos}, \ie the charged and neutral mass eigenstates
corresponding to electroweak gauginos and higgsinos. Charginos
$(\tilde{W}^{\pm}, \tilde{H}^{\pm})$ mix via the $2 \times 2$ matrix
\begin{equation}
\label{chargino}
\left(
\begin{array}{cc}
M_2 & \sqrt{2} m_W \sin \beta \\
\sqrt{2} m_W \cos \beta & \mu
\end{array}
\right),
\end{equation}
whose mass eigenstates are denoted by $\tilde{\chi}^{\pm}_k$ ($k=1,2$),
and neutralinos $(\tilde{B},\tilde{W}_3,\tilde{H}_1^0,\tilde{H}_2^0)$
mix via the $4 \times 4$ matrix
\begin{equation}
\left(
\begin{array}{cccc}
M_1 & 0 & - m_Z \cos \beta \sin \theta_W & m_Z \sin \beta \sin \theta_W \\
0 & M_2 &   m_Z \cos \beta \cos \theta_W & - m_Z \sin \beta \cos \theta_W \\
- m_Z \cos \beta \sin \theta_W & m_Z \cos \beta \cos \theta_W & 0 & - \mu  \\
 m_Z \sin \beta \sin \theta_W & - m_Z \sin \beta \cos \theta_W & - \mu & 0  \\
\end{array}
\right),
\end{equation}
whose mass eigenstates are denoted by $\tilde{\chi}_i^0$ ($i=1,2,3,4$).
Notice that the lightest neutralino $\tilde{\chi} \equiv \tilde{\chi}_1^0$,
which in most of the acceptable parameter space is the LSP, is
in general a non-trivial admixture of gauginos and higgsinos, and not just
a pure photino $\tilde{\gamma} \equiv \cos \theta_W \tilde{B} + \sin
\theta_W \tilde{W}_3$ as often assumed in phenomenological studies. In
the MSSM, all the masses and couplings in the chargino-neutralino sector
can be characterized by the three parameters $m_{1/2}, \mu$, and $\tb$.

To give an idea of the structure of the MSSM $R$-odd spectrum, we show in
figs.~9 and 10 (updated from ref.~[\ref{rrz}]) contours of some selected
sparticle masses in the $(m_0,m_{1/2})$ and in the $(\mu,m_{1/2})$ planes,
respectively, for the representative values $\tb = 2$ and $\tb = 10$.

\subsection{Searches for sleptons}

The most stringent limits on sleptons come from unsuccessful searches
for the processes $Z \to \tilde{l}^+ \tilde{l}^-$ and $Z \to \tilde{\nu}
\tilde{\ov{\nu}}$ at LEP~I. In the mass range of interest, and assuming
that $\tilde{\chi}$ is the LSP, the main decay modes are $\tilde{l}^{\pm}
\to l^{\pm} \tilde{\chi}$ and $\tilde{\nu} \to \nu \tilde{\chi}$. Indirect
but powerful information can be extracted from the precise measurements
of the total and partial $Z$ widths. Direct searches are sensitive to
charged sleptons only, and look for acoplanar lepton pairs with missing
transverse momentum. Experimental details on slepton searches at LEP~I
can be found in refs.~[\ref{sleptons},\ref{lepsearches}]. Crudely
speaking, one can summarize the present limits by $m_{\tilde{l}},
m_{\tilde{\nu}} \simgt m_Z/2$. In the future, LEP~II will be sensitive
to charged sleptons up to $m_{\tilde{l}} \simeq$ 80--90 GeV, whereas the
limits on $m_{\tilde{\nu}}$ are not expected to improve. At large hadron
colliders like the LHC-SSC, slepton searches appear problematic [\ref{lhcsle}],
since the Drell-Yan production cross-sections are small and the
backgrounds are large. It is then clear that high-energy $\epem$ colliders can
play a very important role in slepton searches, as will be now outlined.

Theoretical aspects of slepton production and decay at EE500 have been
recently investigated in ref.~[\ref{vienna}]. The production mechanisms
considered in this study are
\be
\label{selpro}
\epem \to \sell^+ \sell^-, \; \selr^+ \selr^-, \; \sell^{\pm} \selr^{\mp},
\ee
\be
\label{smupro}
\epem \to \smul^+ \smul^-, \; \smur^+ \smur^-,
\ee
\be
\label{snupro}
\epem \to \snu \snubar.
\ee
The first two processes in (\ref{selpro}) occur via $(\gamma,Z)$ exchange in
the s-channel and $\tilde{\chi}^0_i$ exchange in the t-channel. The last
process in (\ref{selpro}) receives only t-channel contributions, the two
processes in (\ref{smupro}) only s-channel contributions. The processes in
(\ref{snupro}) occur via Z exchange in the s-channel, with
$\tilde{\chi}^{\pm}_k$ exchange in the t-channel also contributing in the
case of $\snu_e$. In general, then, the production cross-sections depend
not only on the slepton masses, but also on the parameters of the
chargino-neutralino sector.

As far as decay modes are concerned, one has to take into account the
possibility of cascade decays,
$\tilde{l}^{\pm}_{L,R} \to l^{\pm} \tilde{\chi}^0_{i \ne 1} \to \ldots$,
$\slepl^{\pm}      \to \nu     \tilde{\chi}^{\pm}_k     \to \ldots$,
$\snu              \to \nu     \tilde{\chi}^0_{i \ne 1} \to \ldots$,
$\snu              \to l^{\pm} \tilde{\chi}^{\mp}_k     \to \ldots$,
in addition to the direct decays $\tilde{l}^{\pm}_{L,R} \to l^{\pm}
\tilde{\chi}$, $\snu \to \nu \tilde{\chi}$. Also the relevant branching
ratios depend on the parameters of the chargino-neutralino sector.

A detailed analysis of the whole parameter space will not be attempted here.
The most likely case, in view of the theoretical constraints on the MSSM,
seems to be the one in which the lightest sleptons are $\slepr^{\pm}$
$(l=e,\mu,\tau)$. In this case one obtains [\ref{vienna}] sizeable
cross-sections, $O(10 \; {\rm fb})$ or more, up to slepton masses
of 80--90 $\%$ of the beam energy, which should allow for a relatively
easy detection if the mass difference $(m_{\selr} - m_{\tilde{\chi}})$
is not too small [\ref{grivaz}].

\subsection{Searches for squarks and gluinos}

Being strongly interacting sparticles, squarks and gluinos are best searched
for at hadron colliders. Production cross-sections for $\tilde{g} \tilde{g}$,
$\tilde{g} \tilde{q}$, $\tilde{q} \tilde{\ov{q}}$ pair-production in $pp$
or $p \ov{p}$ collisions are relatively model-independent functions of
$m_{\tilde{g}}$ and $m_{\tilde{q}}$. As far as signatures are concerned,
one has to distinguish two main possibilities: if $\mglu < \msq$, then
$\tilde{q} \to q \tilde{g}$ immediately after production, and the final
state is determined by $\tilde{g}$ decays; if $\msq < \mglu$, then
$\tilde{g} \to \tilde{q} \ov{q}$ immediately after production, and the final
state is determined by $\tilde{q}$ decays. The first case is favoured by
the theoretical constraints of the MSSM. In old experimental analyses, it
was customary to work under a certain set of assumptions: 1) five or six
$(\tilde{q}_L,\tilde{q}_R)$ mass-degenerate squark flavours; 2) LSP
$ \equiv \tilde{\gamma}$, with mass negligible with respect
to $\msq,\mglu$; 3) the dominant decay modes of squarks and gluinos
are the direct ones, $\tilde{g} \to q \ov{q} \tilde{\gamma}$ if $\mglu
< \msq$ and $\tilde{q} \to q \tilde{\gamma}$ if $\msq < \mglu$.
The signals to be looked for are then multijet events with a large amount
of missing transverse momentum. To derive reliable limits, however, one
has to take into account that the above assumptions are in general incorrect.
For example, one can have cascade decays $\tilde{g} \to q \ov{q}
\tilde{\chi}^0_{i \ne 1}, q' \ov{q} \chi^{\pm}_k \to \ldots$ and $\tilde{q}
\to q \tilde{\chi}^0_{i \ne 1}, q' \tilde{\chi}^{\pm}_k \to \ldots$.
The effects of these cascade
decays become more and more important as one moves to higher and higher
squark and gluino masses. Taking all this into account, the present limits
from the Tevatron collider are roughly $\msq \simgt 150 \gev$, $\mglu
\simgt 135 \gev$ [\ref{squarks}]. At the LHC and SSC, one should be
able to explore squark and gluino masses up to 1 TeV and probably more
[\ref{lhcsqu}]. In general, therefore, EE500 will not be competitive for
squark and gluino searches. Its cleaner environment, however, could be
exploited for a detailed study of squark properties if they  are
discovered at sufficiently low mass. Also, there are
special situations which might be difficult to study at large hadron
colliders: for example, the case of a stop squark significantly lighter
than all the other squarks. The threshold behaviour for stop production
in $\epem$ collisions has been recently studied in ref.~[\ref{stop}].

\subsection{Searches for charginos and neutralinos}

The most stringent limits on charginos and neutralinos come
[\ref{charginos},\ref{sleptons},\ref{lepsearches}] from unsuccessful LEP~I
searches for the processes $Z \to \tilde{\chi} \tilde{\chi}$
(contributing to the invisible Z width), $Z \to \tilde{\chi}
\tilde{\chi}^0_{i\ne1}$ (originating spectacular one-sided events)
and $Z \to \tilde{\chi}_1^+ \tilde{\chi}_1^-, \; \tilde{\chi}^0_{i\ne1}
\tilde{\chi}^0_{j\ne1}$ (originating acoplanar leptons or jets accompanied
by missing energy). The presently excluded region of the $(\mu,m_{1/2})$
plane is shown, for the two representative values $\tb=2$ and $\tb = 10$,
in fig.~10. As a crude summary, one could say that all states different
from $\tilde{\chi}$ have to be heavier than $m_Z/2$, whereas LEP data alone
would still allow for
arbitrarily light $\tilde{\chi}$. For LEP~II searches, the most effective
process should be $\tilde{\chi}_1^+ \tilde{\chi}_1^-$ pair production,
with $\tilde{\chi} \tilde{\chi}^0_{i\ne1}$ pair production slightly less
effective in probing parameter space because of the smaller cross-section.
At large hadron colliders [\ref{lhccha}], it seems very difficult to improve
the LEP~II sensitivity significantly, especially if the top quark mass is
significantly smaller than 200 GeV, as now favoured by radiative-correction
analyses.

Theoretical aspects of chargino and neutralino production and decays
at EE500 have been recently studied in ref.~[\ref{vienna}],
and experimental simulations are reported in ref.~[\ref{grivaz}].
In the case of charginos, the most important production
diagrams involve the s-channel exchange of $(\gamma,Z)$ and the t-channel
exchange of $\snu_e$. The cross-section then depends not only on the
parameters of the chargino-neutralino sector, but also on the sneutrino
mass, and there can be significant destructive interference between
the two classes of diagrams. As for chargino decays, if the sneutrino
is light enough the dominant decay mode is $\tilde{\chi}_1^{\pm} \to
l^{\pm} \snu$, whereas in the case of a heavy sneutrino the dominant decay
modes are $\tilde{\chi}^{\pm}_1 \to q \ov{q}' \tilde{\chi}, \; l^{\pm} \nu
\tilde{\chi}$. Experimental analyses show that at EE500
one can enormously extend the parameter space accessible to LEP~II, and
reach chargino masses of the order of 80--90 $\%$ of the beam energy,
provided that the mass difference $m_{\tilde{\chi}^{\pm}_1} -
m_{\tilde{\chi}}$ is not too small: this unfortunate situation could occur
in the region of parameter space where $|\mu| << m_{1/2}$.

\section{Conclusions}

In summary, in this talk we have argued that the MSSM is a
calculable, theoretically motivated and phenomenologically acceptable
extension of the SM. Of course, only experiment can tell if low-energy
supersymmetry is actually realized in Nature, but, to use the words of
one of the speakers at this Workshop, searching for supersymmetry
does not look like fishing in a dead sea. For a global view of the present
limits and of the discovery potential of future machines, including
EE500, it is useful to look again at the most important parameters of
the MSSM
\be
m_A,
\;\;\;\;
\tb \, ,
\;\;\;
m_0 \, ,
\;\;\;
m_{1/2} \, ,
\;\;\;
\mu \, ,
\ee
which, together with $m_t$, determine the main features of its
particle spectrum.

The Higgs sector mainly depends on $(m_A,\tb)$, but also $m_t$ and
(to a lesser extent) the other parameters play a role via the large
radiative corrections. As an example, we have considered the case
$m_t = 140 \gev$, $\msq = 1 \tev$, $A_t=A_b=\mu=0$, summarized in
figs.~3--8. Fig.~3 shows that LEP~I, despite its remarkable achievements,
has explored only a small part, roughly $m_A \simlt 45 \gev$, of
the natural parameter space for the MSSM Higgs bosons. A much greater
sensitivity will be achieved at LEP~II, where, for standard values of
the machine parameters $(\sqrt{s} = 190 \gev, \; \int {\cal L} dt =
500 \pb^{-1})$, one should be able to test $m_A \simlt 80 \gev$,
$\tb \simlt 3$. However, as a result of the large radiative corrections,
the rest of the $(m_A,\tb)$ parameter space will not be accessible to
LEP~II. The LHC-SSC can greatly improve over LEP~II, as shown in fig.~4.
The most promising experimental signatures are $h,H \to \gamma \gamma$
(inclusive or in association with $W$ or $t \ov{t}$), $H \to ZZ \to 4l^{\pm}$,
$H,A \to \tau^+ \tau^-$, $t \to b H^+ \to b \tau^+ \nu_{\tau}$.
Combined, they might be able to probe the whole $(m_A,\tb)$ plane, with
the exception of $m_Z \simlt m_A \simlt 200 \gev$, $2 \simlt \tb \simlt 10$.
For our choice of parameters, then, the overlap between LEP~II and the
LHC-SSC is likely not to be complete, giving rise to a possible violation of
the so-called `no-lose theorem'. A further increase of the LEP~II
energy might save the day. On the other hand, as shown in figs.~5--8,
EE500 is guaranteed to observe at least one neutral Higgs boson or to
rule out the MSSM. In particular, in the region of the parameter space
which is most difficult for the LHC-SSC, EE500 can perform a detailed
spectroscopy of the MSSM Higgs sector, observing {\em all} its physical
states.

As for the ($R$-odd) supersymmetric particles, the situation is summarized
in figs.~9 and 10. Again, we can see that the already impressive limits
obtained by LEP~I and Tevatron have ruled out only a small part of the
natural parameter space. LEP~II and the upgraded Tevatron will provide
higher but still limited sensitivities, corresponding roughly to
$m_{\tilde{l}},m_{\tilde{\chi}^{\pm}} \simlt$ 80--90 GeV, $m_{\tilde{g}},
\msq \simlt 200 \gev$. The LHC-SSC should definitely cover the rest of the
parameter space, via gluino and squark searches up to masses of 1 TeV or
higher. However, a comparable sensitivity can be reached by EE500 via
chargino and slepton searches up to masses of 200 GeV or even higher.
If no signal of supersymmetry is found at the LHC-SSC, EE500 can provide
the definitive confirmation that the MSSM is ruled out, in a cleaner
environment and in a more model-independent way. In the optimistic
case that a signal is found at the LHC-SSC, EE500 would constitute a unique
facility for the direct production of weakly interacting supersymmetric
particles, which should allow for a detailed spectroscopy of the MSSM.

In conclusion, for what concerns supersymmetry, EE500 is the ideal
complement to the LHC-SSC, and the case for it could not be stronger.

\section*{Acknowledgements}
I am grateful to my collaborators A. Brignole, J. Ellis, Z. Kunszt and
G. Ridolfi for their contributions to our common work on supersymetric
Higgs bosons, and to them and F. Pauss for discussions concerning the
present paper.
\newpage
\section*{References}
\begin{enumerate}
\item
\label{fayet}
P. Fayet, Nucl. Phys. B90 (1975) 104,
Phys. Lett. 64B (1976) 159 and 69B (1977) 489.
\item
\label{mssm}
For reviews and references, see, e.g.:
\\
H.-P. Nilles, Phys. Rep. 110 (1984) 1;
\\
H.E. Haber and G.L. Kane, Phys. Rep. 117 (1985) 75;
\\
S. Ferrara, ed., `Supersymmetry' (North-Holland, Amsterdam, 1987);
\\
R. Barbieri, Riv. Nuovo Cimento 11 (1988) 1.
\item
\label{grivaz}
J.-F. Grivaz, these Proceedings.
\item
\label{susy}
Yu.A. Gol'fand and E.P. Likhtman, JETP Lett. 13 (1971) 323;
\\
D.V. Volkov and V.P. Akulov, Phys. Lett. 46B (1973) 109;
\\
J. Wess and B. Zumino, Nucl. Phys. B70 (1974) 39.
\item
\label{sugra}
D.Z. Freedman, P. van Nieuwenhuizen and S. Ferrara,
Phys. Rev. B13 (1976) 3214;
\\
S. Deser and B. Zumino, Phys. Lett. 62B (1976) 335.
\item
\label{hls}
R. Haag, J. Lopuszanski and M. Sohnius, Nucl. Phys. B88 (1975) 257.
\item
\label{gsw}
For reviews and references see, e.g.:
\\
M.B. Green, J.H. Schwarz and E. Witten, `Superstring Theory'
(University Press, Cambridge, 1987);
\\
B. Schellekens, ed., `Superstring construction' (North-Holland,
Amsterdam, 1989).
\item
\label{nat}
K. Wilson, as quoted in L. Susskind, Phys. Rev. D20 (1979) 2619;
\\
E. Gildener and S. Weinberg, Phys. Rev. D13 (1976) 3333;
\\
G. 't Hooft, in `Recent developments in gauge theories', Cargese
Lectures 1979 (Plenum Press, New York, 1980).
\item
\label{lphep}
For reviews and references, see, e.g.:
\\
J.R. Carter, J. Ellis and T. Hebbeker, Rapporteur's talks given at the
LP-HEP '91 Conference, Geneva, 1991, to appear in the Proceedings, and
references therein.
\item
\label{nrt}
J. Wess and B. Zumino, Phys. Lett. B49 (1974) 52;
\\
J. Iliopoulos and B. Zumino, Nucl. Phys. B76 (1974) 310;
\\
S. Ferrara, J. Iliopoulos and B. Zumino, Nucl. Phys. B77 (1974) 413;
\\
M.T. Grisaru, W. Siegel and M. Rocek, Nucl. Phys. B159 (1979) 429;
\\
S. Ferrara, L. Girardello and F. Palumbo, Phys. Rev. D20 (1979) 403.
\item
\label{les}
L. Maiani, Proc. Summer School on Particle Physics, Gif-sur-Yvette, 1979
(IN2P3, Paris, 1980), p. 1;
\\
M. Veltman, Acta Phys. Polon. B12 (1981) 437;
\\
E. Witten, Nucl. Phys. B188 (1981) 513;
\\
see also:
\\
S. Weinberg, Phys. Lett. 82B (1979) 387.
\item
\label{soft}
L. Girardello and M.T. Grisaru, Nucl. Phys. B194 (1982) 65.
\item
\label{fcnc}
J. Ellis and D.V. Nanopoulos, Phys. Lett. B110 (1982) 44;
\\
R. Barbieri and R. Gatto, Phys. Lett. B110 (1982) 211.
\item
\label{cpnew}
Y. Kizukuri and N. Oshimo, contribution to the Proceedings of the
Workshop `$\epem$ Linear Colliders at 500 GeV: the Physics Potential',
Hamburg, 1991, and references therein.
\item
\label{cp}
J. Ellis, S. Ferrara and D.V. Nanopoulos, Phys. Lett. B114 (1982) 231;
\\
W. Buchm\"uller and D. Wyler, Phys. Lett. B121 (1983) 321;
\\
J. Polchinski and M.B. Wise, Phys. Lett. B125 (1983) 393;
\\
F. del Aguila, J.A. Grifols, A. Mendez, D.V. Nanopoulos and M. Srednicki,
Phys. Lett. B129 (1983) 77;
\\
J.-M. Fr\`ere and M.B. Gavela, Phys. Lett. B132 (1983) 107.
\item
\label{singlet}
R.K. Kaul and P. Majumdar, Nucl. Phys. B199 (1982) 36;
\\
R. Barbieri, S. Ferrara and C.A. Savoy, Phys. Lett. B119 (1982) 36;
\\
H.P. Nilles, M. Srednicki and D. Wyler, Phys. Lett. B120 (1983) 346;
\\
J.M. Fr\`ere, D.R.T. Jones and S. Raby, Nucl. Phys. B222 (1983) 11;
\\
J.-P. Derendinger and C. Savoy, Nucl. Phys. B237 (1984) 307;
\\
J. Ellis, J.F. Gunion, H.E. Haber, L. Roszkowski and F. Zwirner,
Phys. Rev. D39 (1989) 844.
\item
\label{muproblem}
J. Polchinski and L. Susskind, Phys. Rev. D26 (1982) 3661;
\\
J.E. Kim and H.-P. Nilles, Phys. Lett. B138 (1984) 150;
\\
L. Hall, J. Lykken and S. Weinberg, Phys. Rev. D27 (1983) 2359;
\\
G. F. Giudice and A. Masiero, Phys. Lett. 206B (1988) 480;
\\
K. Inoue, M. Kawasaki, M. Yamaguchi and T. Yanagida,
preprint TU-373 (1991);
\\
J.E. Kim and H.-P. Nilles, Phys. Lett. B263 (1991) 79;
\\
E.J. Chun, J.E. Kim and H.-P. Nilles, preprint SNUTP-91-25.
\item
\label{instab}
H.-P. Nilles, M. Srednicki and D. Wyler, Phys. Lett. B124 (1983) 337;
\\
A.B. Lahanas, Phys. Lett. B124 (1983) 341;
\\
L. Alvarez-Gaum\'e, J. Polchinski and M.B. Wise, Nucl. Phys. B221 (1983)
495;
\\
A. Sen, Phys. Rev. D30 (1984) 2608 and D32 (1985) 411.
\item
\label{sinnew}
U. Ellwanger and M. Rausch de Traubenberg, contribution to same Proc.
as ref.~[\ref{cpnew}], and references therein;
\\
B.R. Kim, S.K. Oh and A. Stephan, ibid., and references therein.
\item
\label{rbreak}
L.J.  Hall and M. Suzuki, Nucl. Phys. B231 (1984) 419;
\\
F. Zwirner, Phys. Lett. B132 (1983) 103.
\item
\label{spontr}
C. Aulakh and R.N. Mohapatra, Phys. Lett. B119 (1983) 136;
\\
G.G. Ross and J.W.F. Valle, Phys. Lett. B151 (1985) 375;
\\
J. Ellis, G. Gelmini, C. Jarlskog, G.G. Ross and J.W.F. Valle,
Phys. Lett. B150 (1985) 142.
\item
\label{dreiner}
H. Dreiner and S. Lola, contribution to same Proc.
as ref.~[\ref{cpnew}], and references therein.
\item
\label{gcond}
H.P. Nilles, Phys. Lett. 115B (1982) 193 and Nucl. Phys. B217 (1983) 366;
\\
S. Ferrara, L. Girardello and H.P. Nilles, Phys. Lett. 125B (1983) 457;
\\
J.-P. Derendinger, L.E. Ib\'a\~nez and H.P. Nilles, Phys. Lett.
155B (1985) 65;
\\
M. Dine, R. Rohm, N. Seiberg and E. Witten, Phys. Lett. 156B (1985) 55.
\item
\label{ss}
J. Scherk and J.H. Schwarz, Phys. Lett. B82 (1979) 60 and
Nucl. Phys. B153 (1979) 61;
\\
R. Rohm, Nucl. Phys. B237 (1984) 553;
\\
C. Kounnas and M. Porrati, Nucl. Phys. B310 (1988) 355;
\\
S. Ferrara, C. Kounnas, M. Porrati and F. Zwirner, Nucl. Phys.
B318 (1989) 75;
\\
C. Kounnas and B. Rostand, Nucl. Phys. B341 (1990) 641;
\\
I. Antoniadis, Phys. Lett. B246 (1990) 377.
\item
\label{gqw}
H. Georgi, H.R. Quinn and S. Weinberg, Phys. Rev. Lett. 33 (1974) 451.
\item
\label{ms}
W.A. Bardeen, A. Buras, D. Duke and T. Muta, Phys. Rev. D18 (1978) 3998.
\item
\label{gg}
H. Georgi and S.L. Glashow, Phys. Rev. Lett. 32 (1974) 438.
\item
\label{cz}
For reviews and references, see, \eg:
\\
G. Costa and F. Zwirner, Riv. Nuovo Cimento 9 (1986) 1.
\item
\label{costa}
U. Amaldi, A. B\"ohm, L.S. Durkin, P. Langacker, A.K. Mann,
W.J. Marciano, A. Sirlin and H.H. Williams, Phys. Rev. D36 (1987) 1385;
\\
G. Costa, J. Ellis, G.L. Fogli, D.V. Nanopoulos and F. Zwirner,
Nucl. Phys. B297 (1988) 244.
\item
\label{amaldi}
J. Ellis, S. Kelley and D.V. Nanopoulos, Phys. Lett. B249 (1990) 442
and B260 (1991) 131;
\\
P. Langacker and M. Luo, Phys. Rev. D44 (1991) 817;
\\
U. Amaldi, W. de Boer and H. F\"urstenau, Phys.
Lett. B260 (1991) 447;
\\
F. Anselmo, L. Cifarelli, A. Petermann and A. Zichichi,
preprint CERN-PPE/91-123.
\item
\label{drw}
S. Dimopoulos, S. Raby and F. Wilczek, Phys. Rev. D24 (1981) 1681;
\\
L.E. Ib\'a\~nez and G.G. Ross, Phys. Lett. 105B (1981) 439.
\item
\label{grz}
G. Gamberini, G. Ridolfi and F. Zwirner, Nucl. Phys. B331 (1990) 331.
\item
\label{adhoc}
H. Georgi and D.V. Nanopoulos, Nucl. Phys. B159 (1979) 16;
\\
F. del Aguila and L.E. Ib\'a\~nez, Nucl. Phys. B177 (1981) 60;
\\
L.E. Ib\'a\~nez, Nucl. Phys. B181 (1981) 105;
\\
P. Frampton and S.L. Glashow, Phys. Lett. 131B (1983) 340;
\\
U. Amaldi, W. de Boer, P.H. Frampton, H. F\"urstenau and J.T. Liu,
preprint CERN-PPE/91-233.
\item
\label{thr}
S. Weinberg, Phys. Lett. B91 (1980) 51;
\\
L.J. Hall, Nucl. Phys. B178 (1980) 75;
\\
T.J. Goldman and D.A. Ross, Nucl. Phys. B171 (1980) 273;
\\
P. Bin\'etruy and T. Sch\"ucker, Nucl. Phys. B178 (1981) 293, 307;
\\
I. Antoniadis, C. Kounnas and C. Roiesnel, Nucl. Phys. B198 (1982) 317.
\item
\label{twoloop}
M.B. Einhorn and D.R.T. Jones, Nucl. Phys. B196 (1982) 475;
\\
M.E. Machacek and M.T. Vaughn, Nucl. Phys. B236 (1984) 221.
\item
\label{dim5}
S. Weinberg, Phys. Rev. D26 (1982) 287;
\\
N. Sakai and T. Yanagida, Nucl. Phys. B197 (1982) 533.
\item
\label{susy5}
S. Dimopoulos and H. Georgi, Nucl. Phys. B193 (1981) 150;
\\
N. Sakai, Z. Phys. C11 (1982) 153.
\item
\label{bh}
R. Barbieri and L.J. Hall, preprint LBL-31238 (1991);
\\
J. Ellis, S. Kelley and D.V. Nanopoulos, preprint CERN-TH.6140/91;
\\
G.G. Ross and R.G. Roberts, preprint RAL-92-005.
\item
\label{string}
V.S. Kaplunovsky, Nucl. Phys. B307 (1988) 145;
\\
L. Dixon, V.S. Kaplunovsky and J. Louis, Nucl. Phys. B355 (1991) 649;
\\
J.P. Derendinger, S. Ferrara, C. Kounnas and F. Zwirner, preprint
CERN-TH.6004/91, to appear in Nucl. Phys. B;
\\
G. Lopez-Cardoso and B.A. Ovrut, preprint UPR-0464T (1991);
\\
J. Louis, preprint SLAC-PUB-5527 (1991);
\\
I. Antoniadis, K.S. Narain and T. Taylor,
Phys. Lett. B267 (1991) 37;
\\
G. Lopez-Cardoso and B.A. Ovrut, preprint UPR-0481T (1991);
\\
J.P. Derendinger, S. Ferrara, C. Kounnas and F. Zwirner,
Phys. Lett. B271 (1991) 307.
\item
\label{drbar}
W. Siegel, Phys. Lett. B84 (1979) 193;
\\
D.M. Capper, D.R.T. Jones and P. van Nieuwenhuizen, Nucl. Phys.
B167 (1980) 479;
\\
I. Antoniadis, C. Kounnas and K. Tamvakis, Phys. Lett. B119 (1982) 377.
\item
\label{flipped}
I. Antoniadis, J. Ellis, J. Hagelin and D.V. Nanopoulos, Phys. Lett.
B231 (1989) 65, and references therein.
\item
\label{lacaze}
I. Antoniadis, J. Ellis, R. Lacaze and D.V. Nanopoulos,
Phys. Lett. B268 (1991) 188.
\item
\label{structure}
I. Antoniadis, J. Ellis, S. Kelley and D.V. Nanopoulos, preprint
CERN-TH.6169/91;
\\
L.E. Ib\'a\~nez, D. L\"ust and G.G. Ross, Phys. Lett. B272 (1991) 251.
\item
\label{dm}
S. Weinberg, Phys. Rev. Lett. 50 (1983) 387;
\\
H. Goldberg, Phys. Rev. Lett. 50 (1983) 1419;
\\
L.M. Krauss, Nucl. Phys. B227 (1983) 556.
\item
\label{ehnos}
J. Ellis, J.S. Hagelin, D.V. Nanopoulos, K.A. Olive and M. Srednicki,
Nucl. Phys. B238 (1984) 453.
\item
\label{dmnew}
K.A. Olive and M. Srednicki, Nucl. Phys. B355 (1991) 208;
\\
J. Ellis, D.V. Nanopoulos, L. Roszkowski and D.N. Schramm,
Phys. Lett. B245 (1990) 251;
\\
L. Roszkowski, Phys. Lett. B262 (1991) 59;
\\
J.L. Lopez, K. Yuan and D.V. Nanopoulos, Phys. Lett.
B267 (1991) 219;
\\
J. Ellis and L. Roszkowski, preprint CERN-TH.6260/91, UM-TH-91-25.
\item
\label{susyrad}
T.K. Kuo and N. Nakagawa, Nuovo Cimento Lett. 36 (1983) 560;
\\
L. Alvarez-Gaum\'e, J. Polchinski and M.B. Wise, Nucl. Phys. B221 (1983) 495;
\\
R. Barbieri and L. Maiani, Nucl. Phys. B224 (1983) 32;
\\
C.S. Lim, T. Inami and N. Sakai, Phys. Rev. D29 (1984) 1488;
\\
E. Eliasson, Phys. Lett. B147 (1984) 65;
\\
Z. Hioki, Progr. Theor. Phys. 73 (1985) 1283;
\\
J.A. Grifols and J. Sol\`a, Phys. Lett. B137 (1984) 257 and Nucl. Phys.
B253 (1985) 47;
\\
R. Barbieri, M. Frigeni, F. Giuliani and H.E. Haber, Nucl. Phys. B341
(1990)  309;
\\
A. Bilal, J. Ellis and G.L. Fogli, Phys. Lett. B246 (1990) 459;
\\
A. Djouadi, G. Girardi, C. Verzegnassi, W. Hollik and F.M. Renard,
Nucl. Phys. B249 (1991) 48;
\\
M. Drees and K. Hagiwara, Phys. Rev. D42 (1990) 1709;
\\
M. Boulware and D. Finnell, Phys. Rev. D44 (1991) 2054;
\\
M. Drees, K. Hagiwara and A. Yamada, preprint DTP/91/34.
\item
\label{technirad}
M.E. Peskin and T. Takeuchi, Phys. Rev. Lett. 65 (1990) 964;
\\
M. Golden and L. Randall, Nucl. Phys. B361 (1991) 3;
\\
B. Holdom and J. Terning, Phys. Lett. B247 (1990) 88.
\item
\label{guido}
G. Altarelli, preprint CERN-TH.6245/91, talk given
at the LP-HEP '91 Conference, Geneva, July 1991, to
appear in the Proceedings, and references therein.
\item
\label{rge}
R. Barbieri, S. Ferrara, L. Maiani, F. Palumbo and C.A. Savoy, Phys.
Lett. B115 (1982) 212;
\\
K. Inoue, A. Kakuto, H. Komatsu and S. Takeshita, Progr. Theor. Phys. 68
(1982) 927 and 71 (1984) 413.
\item
\label{bagger}
J. Bagger, S. Dimopoulos and E. Mass\'o, Phys. Rev. Lett. 55 (1985) 920.
\item
\label{pendleton}
B. Pendleton and G.G. Ross, Phys. Lett. B98 (1981) 291;
\\
C. Hill, Phys. Rev. D24 (1981) 691.
\item
\label{radbre}
L.E. Ib\'a\~nez and G.G. Ross, Phys. Lett. 110B (1982) 215;
\\
K. Inoue, A. Kakuto, H. Komatsu and S. Takeshita, as in ref.~[\ref{rge}].
\\
L. Alvarez-Gaum\'e, M. Claudson and M.B. Wise, Nucl. Phys. B207 (1982) 96;
\\
J. Ellis, D.V. Nanopoulos and K. Tamvakis, Phys. Lett. B121 (1983) 123.
\item
\label{cw}
S. Coleman and E. Weinberg, Phys. Rev. D7 (1973) 1888;
\\
S. Weinberg, Phys. Rev. D7 (1973) 2887.
\item
\label{hunter}
For reviews and references see, e.g.:
\\
J.F. Gunion, H.E. Haber, G.L. Kane and S. Dawson, `The Higgs Hunter's
Guide' (Addison-Wesley, New York, 1990).
\item
\label{jap1}
Y. Okada, M. Yamaguchi and T. Yanagida,
Prog. Theor. Phys. Lett. 85 (1991) 1.
\item
\label{erz1}
J. Ellis, G. Ridolfi and F. Zwirner, Phys. Lett. B257 (1991) 83.
\item
\label{hh}
H.E. Haber and R. Hempfling, Phys. Rev. Lett. 66 (1991) 1815.
\item
\label{lisher}
S.P. Li and M. Sher, Phys. Lett. B140 (1984) 339.
\item
\label{gt}
J.F. Gunion and A. Turski, Phys. Rev. D39 (1989) 2701
and D40 (1989) 2333.
\item
\label{berger}
M. Berger, Phys. Rev. D41 (1990) 225.
\item
\label{bfc}
R. Barbieri, M. Frigeni and M. Caravaglios, Phys. Lett. B258 (1991) 167.
\item
\label{jap2}
Y. Okada, M. Yamaguchi and T. Yanagida, Phys. Lett. B262 (1991) 54.
\item
\label{bf}
R. Barbieri and M. Frigeni, Phys. Lett. B258 (1991) 395.
\item
\label{erz2}
J. Ellis, G. Ridolfi and F. Zwirner, Phys. Lett. B262 (1991) 477.
\item
\label{yam}
A. Yamada, Phys. Lett. B263 (1991) 233.
\item
\label{eq}
J.R. Espinosa and M. Quir\'os, Phys. Lett. B266 (1991) 389.
\item
\label{berz}
A. Brignole, J. Ellis, G. Ridolfi and F. Zwirner,
Phys. Lett. B271 (1991) 123.
\item
\label{cpr}
P.H. Chankowski, S. Pokorski and J. Rosiek,
Phys. Lett. B274 (1992) 191.
\item
\label{b}
A. Brignole, preprint DFPD/91/TH/28 (1991), to appear in Phys. Lett. B.
\item
\label{drees}
M. Drees and M.N. Nojiri, preprint KEK-TH-305 (1991).
\item
\label{pierce}
D.M. Pierce, A. Papadopoulos and S. Johnson, preprint LBL-31416,
UCB-PTH-91/58.
\item
\label{kznew}
Z. Kunszt and F. Zwirner, preprint CERN-TH.6150/91, ETH-TH/91-7.
\item
\label{clones}
J.F. Gunion, R. Bork, H.E. Haber and A. Seiden, preprint UCD-91-29,
SCIPP-91/34;
\\
H. Baer, M. Bisset, C. Kao and X. Tata, preprint FSU-HEP-911104, UH-511-732-91;
\\
J.F. Gunion and L.H. Orr, preprint UCD-91-15;
\\
V. Barger, M.S. Berger, A.L. Stange and R.J.N. Phillips,
preprint MAD-PH-680 (1991).
\item
\label{b2}
A. Brignole, preprint CERN-TH.6366/92.
\item
\label{lep}
D. D\'ecamp et al. (ALEPH Collaboration), Phys. Lett. B265 (1991) 475;
\\
P. Igo-Kemenes (OPAL Collaboration),
L. Barone (L3 Collaboration),
W. Ruhlmann (DELPHI Collaboration),
talks given at the LP-HEP '91
Conference, Geneva, 1991, to appear in the Proceedings;
\\
M. Davier, Rapporteur's talk at the same Conference, and references therein.
\item
\label{treille}
D. Treille, private communication;
\\
C. Rubbia, Rapporteur's talk given at the LP-HEP '91 Conference, Geneva,
1991, to appear in the Proceedings;
\\
U. Amaldi, these Proceedings.
\item
\label{kz}
Z. Kunszt and F. Zwirner, Proc. Large Hadron Collider Workshop,
Aachen, 1990 (G. Jarlskog and D. Rein, eds.) (CERN 90-10,
ECFA 90-133, Geneva, 1990), Vol.II, p. 578.
\item
\label{pauss}
F. Pauss, lectures given in the CERN Academic Training Programme,
December 1991, and references therein.
\item
\label{ee500}
A. Brignole, J. Ellis, J.F. Gunion, M. Guzzo, F. Olness, G. Ridolfi,
L. Roszkowski and F.~Zwirner, contribution to the Workshop `$\epem$
Linear Colliders at $500 \gev$: the Physics Potential', Hamburg,
1991, to appear in the Proceedings.
\item
\label{haber}
H.E. Haber, these Proceedings.
\item
\label{janot}
P. Janot, contribution to the same Workshop as ref.~[\ref{ee500}].
\item
\label{koma}
S. Komamiya, these Proceedings.
\item
\label{amp}
D.R.T. Jones and S.T. Petcov, Phys. Lett. B84 (1979) 440;
\\
K. Hikasa, Phys. Lett. B164 (1985) 385;
\\
R. Cahn, Nucl. Phys. B255 (1985) 341;
\\
G. Altarelli, B. Mele and F. Pitolli, Nucl. Phys. B287 (1987) 285.
\item
\label{rrz}
G. Ridolfi, G.G. Ross and F. Zwirner, same Proc. as ref.~[\ref{kz}],
Vol.II, p. 605.
\item
\label{sleptons}
D. D\'ecamp et al. (ALEPH Collaboration), Phys. Lett. B236 (1990) 86;
\\
P. Abreu et al. (DELPHI Collaboration), Phys. Lett. B247 (1990) 157;
\\
B. Adeva et al. (L3 Collaboration), Phys. Lett. B233 (1989) 530;
\\
M.Z. Akrawy et al. (OPAL Collaboration), Phys. Lett. B240 (1990) 261.
\item
\label{lepsearches}
D. D\'ecamp et al. (ALEPH Collaboration), preprint CERN-PPE/91-149,
submitted to Physics Reports.
\item
\label{lhcsle}
F. del Aguila, L. Ametller amd M. Quir\'os,  same Proc. as ref.~[\ref{kz}],
Vol.II, p. 663;
\\
F. del Aguila and L. Ametller, Phys. Lett. B261 (1991) 326.
\item
\label{vienna}
A. Bartl, W. Majerotto and B. M\"osslacher, contribution to the same
Workshop as ref.~[\ref{ee500}].
\item
\label{squarks}
H. Baer, X. Tata and J. Woodside, Phys. Rev. D44 (1991) 207.
\item
\label{lhcsqu}
C. Albajar, C. Fuglesang, S. Hellman, E. Nagy, F. Pauss, G. Polesello
and P. Sphicas, same Proc. as ref.~[\ref{kz}],
Vol.II, p. 621;
\\
H. Baer et al., preprint FSU-HEP-901110, to be published in the
Proceedings of the 1990 DPF Summer Study on High Energy Physics,
Snowmass, CO, June 25-July 13, 1990.
\item
\label{stop}
I.I. Bigi, V.S. Fadin and V. Khoze, preprint UND-HEP-91-BIG03.
\item
\label{charginos}
D. D\'ecamp et al. (ALEPH Collaboration), Phys. Lett. B244 (1990) 541;
\\
M.Z. Akrawy et al. (OPAL Collaboration), Phys. Lett. B248 (1990) 211.
\item
\label{lhccha}
R. Barbieri, F. Caravaglios, M. Frigeni and M. Mangano,
same Proc. as ref.~[\ref{kz}], Vol.II, p. 658;
Nucl. Phys. B367 (1991) 28.
\end{enumerate}
\newpage
\section*{Figure captions}
\begin{itemize}
\item[Fig.1:]
Contours of $m_h^{max}$ (the maximum value of $m_h$, reached for $m_A \to
\infty$) in the $(m_t,\tan \beta)$ plane, for $\msq = 1 \tev$.
\item[Fig.2:]
Contours of a) $m_h$ and b) $m_H$, in the $(m_A,\tan \beta)$ plane,
for $\msq = 1 \tev$ and $m_t = 140 \gev$.
\item[Fig.3:]
Schematic representation of the present LEP~I limits and of the future
LEP~II sensitivity in the $(m_A,\tan \beta)$ plane, for $\msq = 1 \tev$
and $m_t = 140 \gev$. The solid lines correspond to the present LEP~I
limits. The dashed lines correspond to $\sigma (e^+ e^- \to
hZ, HZ, hA, HA) = 0.2 \, {\rm pb}$ at $\sqrt{s} = 175 \gev$, which
could be seen as a rather conservative estimate of the LEP~II
sensitivity. The dashed-dotted lines correspond to $\sigma (e^+ e^- \to
hZ, HZ, hA, HA) = 0.05 \, {\rm pb}$ at $\sqrt{s} = 190 \gev$, which
could be seen as a rather optimistic estimate of the LEP~II
sensitivity.
\item[Fig.4:]
Pictorial representation of the future LHC sensitivity in the
$(m_A,\tan \beta)$ plane, for $\msq = 1 \tev$ and $m_t = 140 \gev$.
\item[Fig.5:]
Contours of a) $\sigma ( e^+ e^- \rightarrow h Z)$ and b) $\sigma ( e^+
e^- \rightarrow H Z)$, in the $(m_A, \tan \beta)$ plane, for $\sqrt{s}
= 500 \gev$.
\item[Fig.6:]
Contours of a) $\sigma ( e^+ e^- \rightarrow h A)$ and b) $\sigma ( e^+
e^- \rightarrow H A)$, in the $(m_A, \tan \beta)$ plane, for $\sqrt{s} =
500 \gev$.
\item[Fig.7:]
Contours of a) $\sigma ( e^+ e^- \rightarrow h \nu \overline{\nu})$
and b) $\sigma ( e^+ e^- \rightarrow H \nu \overline{\nu})$, in the $(m_A,
\tan \beta)$ plane, for $\sqrt{s} = 500 \gev$.
\item[Fig.8:]
Contours of a) $\sigma ( e^+ e^- \rightarrow h e^+ e^-)$ and
b) $\sigma ( e^+ e^- \rightarrow H e^+ e^-)$, via $ZZ$-fusion,
in the $(m_A, \tan \beta)$ plane, for $\sqrt{s} = 500 \gev$.
\item[Fig.9:]
Present limits and future sensitivity in the $(m_0,m_{1/2})$ plane,
for the representative values a) $\tb =2$, b) $\tb =10$ and using
$\mu$-independent constraints.
The shaded area is excluded by the present data, whereas the solid
lines correspond to the estimated discovery potential of the complete
LEP and Tevatron programs. Dashed lines correspond to fixed values
of an `average' squark mass, defined by the relation $m_{\tilde{q}} =
\sqrt{m_0^2 + 5.5 m_{1/2}^2}$. Dotted lines correspond to fixed values
of the mass of the lightest charged slepton ($\tilde{e^c}$), as given
in the text. The values of the gluino mass as given by eq. (\ref{gausol})
are also shown.
\item[Fig.10:]
Present limits and future sensitivity in the $(\mu,m_{1/2})$ plane,
for the representative values a) $\tb =2$, b) $\tb =10$ and using
$m_0$-independent constraints.
The shaded area is excluded by the present data, whereas the solid
lines correspond to the estimated discovery potential of the complete
LEP and Tevatron programs.
Dashed and dotted lines correspond to
fixed values of the lighest chargino and neutralino mass, respectively.
The values of the gluino mass as given by eq. (\ref{gausol}) are also shown.
\end{itemize}
\end{document}